\documentclass[11pt]{article}

\usepackage{amsmath}
\usepackage{unit}
\usepackage{amssymb}
\usepackage{mathrsfs}
\usepackage[dvipdfmx]{graphicx}
\usepackage{setspace}
\usepackage{verbatim}
\usepackage{url}
\usepackage{dcolumn}
\usepackage{color}
\usepackage{here}
\usepackage{subcaption}
\usepackage{amsthm}
\usepackage{bm}
\usepackage{amscd}
\theoremstyle{definition}

\title{%
Autonomous Vehicle-to-Grid Design for Provision of Frequency Control Ancillary Service and Distribution Voltage Regulation\thanks{This work was supported in part by Japan Science and Technology Agency, Core Research for Evolutional Science and Technology (JST-CREST) Program \#JP-MJCR15K3. (Corresponding author: Yoshihiko Susuki, {\tt susuki@ieee.org})}
}%
\author{%
Shota~Yumiki\footnote{During this work, S. Yumiki, Y. Susuki, and A. Ishigame were with Department of Electrical and Information Systems, Osaka Prefecture University, Japan.}, 
Yoshihiko~Susuki,  
Yuta Oshikubo\footnote{During this work, Y. Oshikubo and Y. Ota were with Department of Electrical and Electronics Engineering, Tokyo City University, Japan. 
Y. Ota is presently with Mobility Systems Design Joint Research Laboratory, Osaka University, Japan.}, 
Yutaka~Ota, 
Ryo~Masegi\footnote{During this work, R. Masegi, A. Kawashima. S. Inagaki, and T. Suzuki were with Department of Mechanical Science and Engineering, Nagoya University, Japan. 
S. Inagaki is presently with Department of Mechatronics, Nanzan University, Japan.},\\ 
Akihiko~Kawashima, 
Atsushi~Ishigame, 
Shinkichi~Inagaki, and 
Tatsuya~Suzuki
}%

\date{}

\begin{document}
\maketitle

\begin{abstract}
We develop a system-level design for the provision of Ancillary Service (AS) for control of electric power grids by in-vehicle batteries, suitably applied to Electric Vehicles (EVs) operated in a sharing service. 
An architecture for cooperation between transportation and energy management systems is introduced that enables us to design an autonomous Vehicle-to-Grid (V2G) for the provision of multi-objective AS: primary frequency control in a transmission grid and voltage amplitude regulation in a distribution grid connected to EVs.  
The design is based on the ordinary differential equation model of distribution voltage, which has been recently introduced as a new physics-based  model, and is utilized in this paper for assessing and regulating the impact of spatiotemporal charging/charging of a large population of EVs to a distribution grid. 
Effectiveness of the autonomous V2G design is evaluated with numerical simulations of realistic models for transmission and distribution grids with synthetic operation data on EVs in a sharing service.   
In addition, we present a hardware-in-the-loop test for evaluating its feasibility in a situation where inevitable latency is involved due to power, control, and communication equipments. 
\end{abstract}

%%%%
%%%%
\section{Introduction}

The coordinate use of batteries in electric vehicles (EVs) has attracted a lot of interest for the control of power grids. 
In a large-scale power transmission grid, a large population of in-vehicle batteries is investigated for the provision of ancillary service (AS) to the so-called transmission system operator (TSO), coined in \cite{Kempton:2005,Tomic} as the vehicle-to-grid (V2G). 
It aims to shift the peak load (called valley filling) and to provide reserves for primary, secondary, and tertiary frequency controls of the transmission grid: see, e.g., \cite{Galus:2013,Hu:2016} for technological aspects and \cite{Gschwendtner:2021} for recent implementation.   
The V2G for the reserve of primary frequency control (PFC) is called frequency response in PJM\footnote{Abbreviation of a regional transmission organization that coordinates the movement of wholesale electricity in all part of 13 states and the District of Columbia in United States of America} \cite{Rebours} and also called frequency-controlled normal operation reserve in the Nordic energy region \cite{FNR}. 
It is required to achieve fast responsiveness of several seconds to several minutes \cite{Kempton:2005}. 
In-vehicle batteries are capable of responding faster than synchronous generators used in large thermal power plants and are hence suitable for the PFC provision. 
Much volume of literature exists for the provision of PFC reserve by in-vehicle batteries: as papers including experiments, see \cite{Ota:2015,Marinelli:2016,Knezovic:2017,Arias:2018}.  
On the other hand, in a small-scale distribution grid, the so-called distribution system operator (DSO) \cite{DSO2} is investigated for procuring AS from EVs \cite{Knezovic:2017b,Arias:2019}. 
The AS is intended for compensating with in-vehicle batteries some of services provided by DSO including congestion prevention and voltage magnitude regulation (see \cite{Knezovic:2017b} for details).  
The coordinate use of in-vehicle batteries for DSO has been reported recently, see references in the comprehensive review \cite{Arias:2019}. 
The situation where a large population of EVs is connected to the distribution grid has motivated a research direction in power grids  for past ten years \cite{Ipakchi}. 

One major concern in the situation is with how the impact of EV's charging/discharging to the distribution voltage is assessed and regulated. 
An EV can move anywhere and thus conduct charging and discharging anywhere (precisely, any location such as households and charging stations) in the distribution grid. 
The charging of EVs uncoordinated in the grid can lead to larger voltage variations and lower power quality. 
As one of the early studies, the authors of \cite{Clement1, Clement2} studied the impact of EV's charging to the distribution voltage based on the power-flow equation and proposed an optimization-based scheduling of EV's charging considering the voltage impact. 
The authors of \cite{Yilmaz:2013} presented a review of the status and implementation of V2G technologies on distributed grids until 2012. 
The assessment and regulation problems are now still actively studied by many groups of researchers, see, e.g., \cite{Wang:2019,Dixon:2020} and references in \cite{Arias:2019}. 
As stated in \cite{Arias:2019}, since EVs are mainly connected to the distribution grid, the provision of AS to TSO may affect the distribution grid, which has been scarcely studied until now. 
One early study is found in \cite{Tokudome:2009} where the PFC and voltage regulation in a small power grid are simultaneously achieved with in-vehicle batteries. 
To the best of our survey, there is no comprehensive research on methodology and tools for managing the charging/discharging of EVs in which they provide the PFC reserve while its physical impact on the distribution voltage is regulated. 
The research is of technological significance since EVs have been becoming more and more popular as new service providers in recent years \cite{Arias:2019}. 

Our research team has developed theory and algorithm for solving this problem by means of EVs in a \emph{sharing service}  \cite{Kawashima:CCTA17,Mizuta:2017,Mizuta:2018,Mizuta:2019,Suzuki}. 
An EV-sharing system has a function of tracking trajectories of EVs to allocate them upon user's requests \cite{Fairley,Bignami:2017} and has attracted interest as the last mile transportation \cite{Tan}. 
It is capable of monitoring and managing the status of in-vehicle batteries such as state-of-charge and degree-of-health, and hence it can work for their effectiveness use for not only its primary concern of transportation but also the AS to exiting power grids.  
The work in \cite{Kawashima:CCTA17}, which motivates the present paper, shows a possibility of cooperative transportation-energy management, in which EV-sharing system and EMS work consistently. 
A computational method was developed in \cite{Mizuta:2017,Mizuta:2018,Mizuta:2019} for synthesizing a spatial pattern of EV operation in terms of charging/discharging modes, which was referred to as the synthesis of \emph{charging/discharging pattern}. 
The method builds upon a new physics-based model of distribution voltage profile, called the nonlinear ordinary differential equation (ODE) model, which is originally derived in \cite{Chertkov:2011} and utilized in \cite{Susuki2,Suzuki} for the impact assessment of shared EVs in a distribution grid. 
The method in \cite{Mizuta:2017,Mizuta:2018,Mizuta:2019} determines the amount of charging/discharging power of \emph{local} charging stations, where individual EVs are connected to a distribution grid, so that the \emph{global} objectives of AS for frequency control and of voltage regulation are achieved. 
This idea is called in \cite{Mizuta:2019} as the provision of \emph{multi-objective AS}\footnote{This implies the achievement of multiple control objectives by provision of a single AS and does not imply the use of any technique of multi-objective optimization in the computational method.} by in-vehicle batteries. 
Here, it should be mentioned that independently from us, the authors of \cite{Iacobucci:2018,Iacobucci:2019} investigated the use of automated, shared EVs for the V2G and stated in \cite{Iacobucci:2018} ``This allows for a direct connection to the high voltage electricity transmission in designated points without overloading the low-voltage distribution network." 
Similar arguments can be found in the recent review \cite{Arias:2019} and the recent paper \cite{Chen:2020}. 
These indicate that the cooperative transportation-energy management, possibly combined with the emergent automated driving, is worth pursuing. 

The cooperative management was conceptually reported in \cite{Kawashima:CCTA17}, but its detailed architecture including a connection to the multi-objective AS and its control system was not developed in \cite{Mizuta:2017,Mizuta:2018,Mizuta:2019,Suzuki}. 
We devote the present paper to show what type of architecture is possible in the cooperation of EV-sharing operator and DSO, and how it enables the provision of multi-objective AS. 
The cooperation implies the exchange of information between the two operators so that they simultaneously achieve multiple functions in the EV sharing and distribution grid: see Section~\ref{sec:framework}. 
The architecture in this paper aims to provide the multi-objective AS---PFC reserve for a transmission grid and voltage amplitude regulation support 
for a distribution grid.
For this, we propose an autonomous V2G design in the architecture based on the preceding papers \cite{Ota:2012,Yumiki}. 
The authors of \cite{Ota:2012} proposed an autonomous scheme of the provision of PFC reserve by distributed EVs, in which its impact on distribution voltage is not considered. 
In \cite{Yumiki}, a computational technique was developed for determining an upper bound for the synthesis of charging/discharging patterns in terms of the voltage regulation. 
The technique is based on the nonlinear ODE model in the similar manner as in \cite{Mizuta:2017,Mizuta:2018,Mizuta:2019}.   
In the design proposed here, charging or discharging power by EVs distributed for multiple stations is regulated with the \emph{local} measurement of frequency for the provision of PFC reserve, while the maximum amount of charging or discharging power is determined \emph{globally} for the voltage regulation. 
Effectiveness of the design is evaluated with numerical simulations of realistic models for transmission and distribution grids with synthetic operation data on EVs in a sharing service.  
In addition to the effectiveness, we use the hardware-in-the loop (HIL) testbed partly developed in \cite{Mizuta:2019} in order to evaluate practical feasibility of the design. 

The main contributions of this paper are three-fold: 
\begin{itemize}
\item The architecture for the cooperation between EV-sharing operator and DSO is introduced (see Section~\ref{sec:framework}).  
This sort of architecture is originally reported in \cite{Galus:2013,Knezovic:2017b} where the cooperation is addressed between DSO and PEV (plug-in EV) aggregator, which plays a key role in the so-called direct control \cite{Galus:2013} that does not actively involve vehicle owners in the control actions imposed to PEVs.   
Our novelty in this paper is therefore to introduce as an PEV aggregator the EV-sharing manager which can monitor and manage the status of in-vehicle batteries in a centralized manner. 
Note that the architecture introduced in this paper is an extended version of \cite{Yumiki} in which no AS requirement was introduced. 
\item The autonomous V2G design considering the voltage regulation is introduced and proven to be effective with numerical simulations. 
The mechanism for the PFC provision is based on the so-called droop-type characteristic \cite{Tokudome:2009,Lopes:2011,Ota:2012,Liu:2013} and is therefore not new. 
In this paper, by combing the mechanism with the technique in \cite{Yumiki}, we propose the design for providing the PFC reserve by in-vehicle batteries in a decentralized manner while regulating (precisely, guaranteeing by design) their impact of to the distribution voltage. 
This design is novel to the best of our survey. 
This contribution is not limited to EVs and applicable to general battery devices connected to distribution grids, which state-of-art is presented in literature, e.g, \cite{Ben-Idris_IEEE-EM}. 
\item The practical feasibility of the autonomous V2G design is established with the Power-HIL testbed. 
The Power-HIL is utilized for testing the coupling of transportation and energy systems, especially, the V2G: see, e.g., \cite{Ota:2015,Marinelli:2016,Knezovic:2017,Martinenas:2017}. 
The importance of HIL for validating the AS provision of power-electronics-interfaced distributed generations like EV batteries is discussed in \cite{Kotsampopoulos}. 
Our Power-HIL testing shows that the dynamics of frequency and voltage under the autonomous V2G design are consistently simulated. 
The simulation is done in a connection of multiple components occurring in practice, such as hardware including EV batteries, software (digital simulator), communication lines, and measurement devices. 
It is then shown that the autonomous V2G design works in a practical situation where inevitable latency due to physical, control, and communication equipments is involved. 
Establishing the feasibility of the design is novel. 
\end{itemize}
It should be noted that the conference proceeding \cite{Yumiki} as a preliminary work of this paper does not contain the autonomous V2G design in Sections~\ref{sec:autonomou} to \ref{sec:results}, which is the main contribution of the present paper. 

The rest of this paper is organized as follows. 
In Section~\ref{sec:framework} we propose the architecture for the cooperation between EV-sharing operator and DSO. 
Section~\ref{sec:ODE} introduces the nonlinear ODE model of distribution voltage, and Section~\ref{sec:main} reviews its use for the computation of upper bound reported in \cite{Yumiki}. 
In Section~\ref{sec:autonomou}, we describe the autonomous V2G design for the provision of multi-objective AS that works in the proposed architecture.  
Effectiveness of the proposed design is evaluated in Section~\ref{sec:exp}. 
Its feasibility testing is presented in Sections~\ref{sec:HIL} and \ref{sec:results}. 
Section~\ref{sec:conclusion} is the conclusion of this paper with a brief summary and future directions.

%%%%
%%%%
\section{Overview of Cooperative Transportation-Energy Management} 
\label{sec:framework}

\begin{figure*}[t]
\centering
\includegraphics[width=.95\hsize]{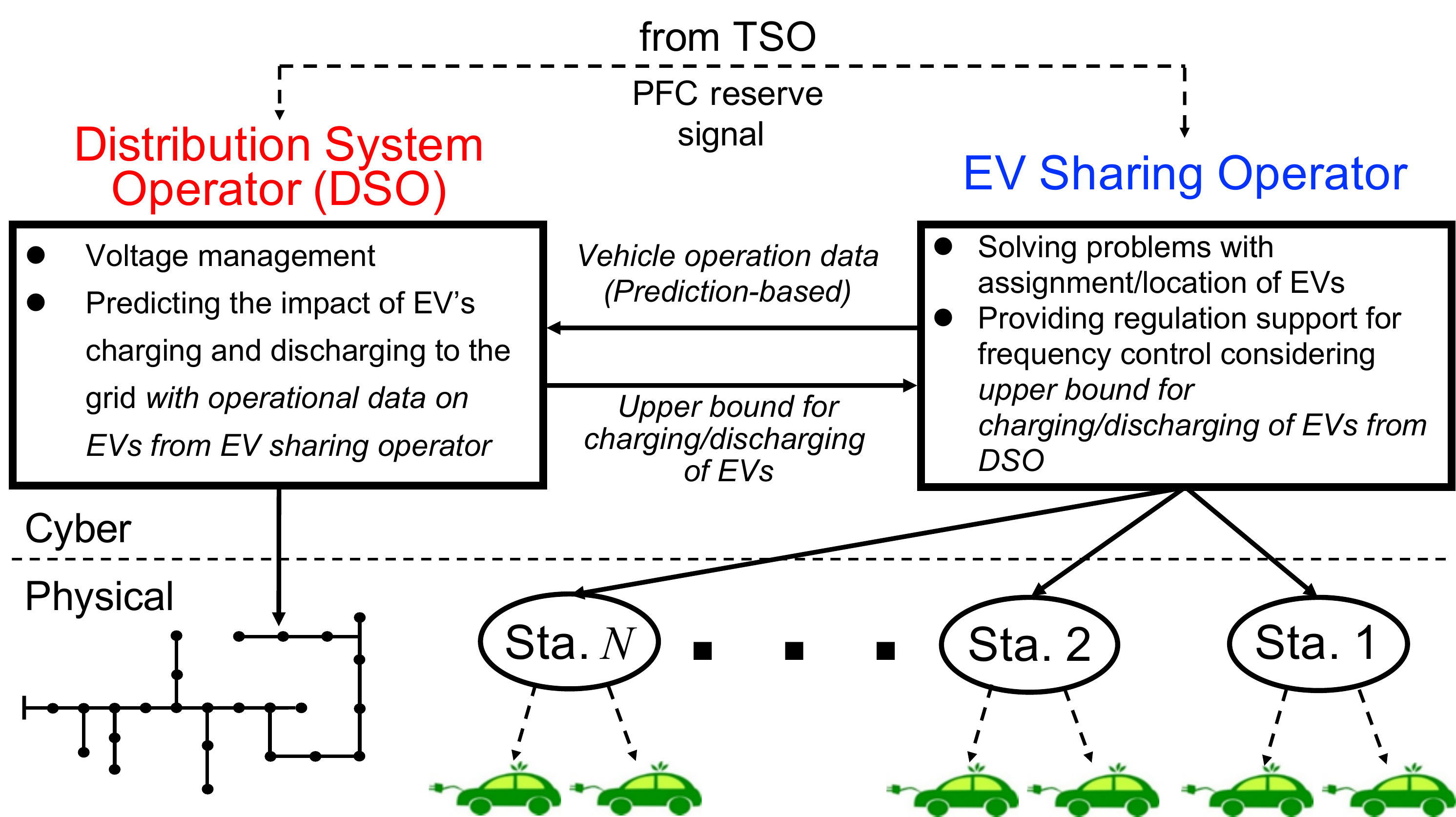}
\vskip 4mm
\caption{% 
Proposed architecture for cooperation between electric vehicle (EV)-sharing operator and distribution system operator (DSO), on which an autonomous vehicle-to-grid (V2G) is designed for the provision of multi-objective ancillary service (AS). 
}%
\label{fig:system}
\end{figure*}

This section introduces an architecture for the cooperation between EV-sharing operator and DSO, as shown in Figure~\ref{fig:system}, on which the autonomous V2G design works.   
The basic function of EV-sharing operator is based on \cite{Masegi:2018}. 
The operator receives operation data of users (including reservations), vehicles, and charging stations. 
Based on the data, the EV-sharing operator determines the assignment of vehicles to users, their reallocation among stations in the system. 
Moreover, it sends to DSO the prediction data of how the number of EVs is changed in time at each charging station.  
In our study, the data include the number of EVs that can contribute to the PFC reserve in AS. 
Using the prediction data, DSO manages a distribution grid spanned in a distinct where the shared vehicles are operated. 
The basic function of DSO is discussed in \cite{DSO2,Arias:2019}.  
According to the references, DSO is responsible for realizing a reliable grid operation, keeping the quality of electricity, minimizing the loss of energy, and responding to normal and emergency situations correctly.
In our study, by using the prediction data, DSO takes a new function of predicting how the charging and discharging of EVs affect the grid. 
In addition to this, DSO utilizes the method proposed in Section~\ref{sec:main} and determines an upper bound for the synthesis of charging/discharging patterns in order to regulate the voltage impact. 
The determined bound is sent from DSO to EV-sharing operator as shown in the upper part of Figure~\ref{fig:system}. 
With the cooperation, we design the autonomous V2G scheme for the shared vehicles in order to provide the PFC reserve to an upper TSO. 
TSO has a function of keeping the balance of demand and supply of electric power in the grid \cite{Rebours}. 
EV-sharing operator sends the upper bound to each station, and then the station manages the charging or discharging of multiple EVs connected there (see the lower part of Figure~\ref{fig:system}) such that the total amount of charged or discharged power is within the receiving upper bound. 
Thereby, the charging and discharging of power can provide the PFC reserve to TSO, while their impact to the distribution voltage can be regulated (precisely, they can support the voltage regulation by DSO). 
Throughout the paper, we will contend that the exchange of information on EV-sharing operator and DSO in the upper part of Figure~\ref{fig:system} is a key enabler for the provision of multi-objective AS: PFC reserve and voltage regulation. 

%%%%
%%%%
\section{ODE-Based Analysis of Distribution Voltage Profile}
\label{sec:review}

%%%%
\subsection{ODE Model} 
\label{sec:ODE}

\begin{figure}[t]
\centering
\includegraphics[width=.7\textwidth]{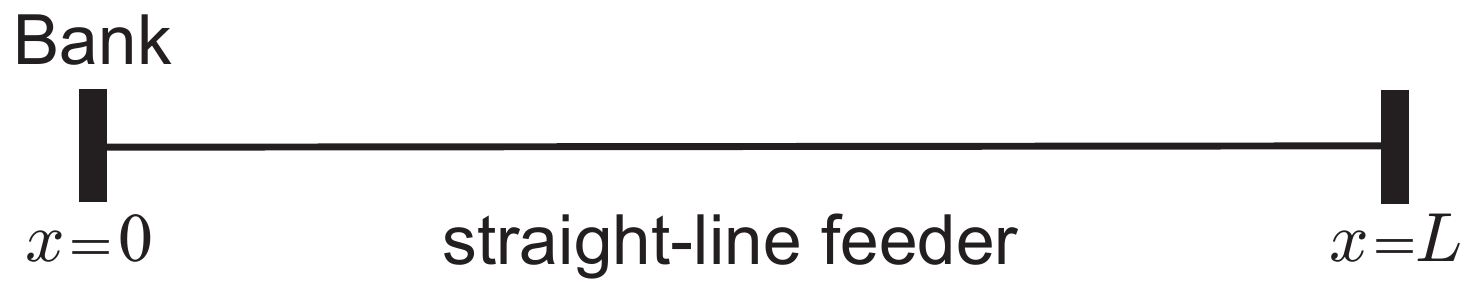}
\caption{%
Single-line representation of balanced three-phase distribution feeder that starts at a bank (transformer), through a finite-length line with length $L$, and ends at a non-loading terminal
}%
\label{fig:feeder1}
\end{figure}

This section introduces the ODE model of distribution voltage profile based on \cite{Chertkov:2011}. 
A single feeder model is shown in Figure~\ref{fig:feeder1} that is straight-line and starts at a bank (transformer), where we introduce the origin of the displacement (location) $x\in\mathbb{R}$ as $x=0$. 
The voltage phasor at the location $x$ is represented by $v(x)\exp\{\ii\theta(x)\}$ where $\ii$ is the imaginary unit, $v(x)$ the \emph{voltage amplitude} in volt [V], and $\theta(x)$ the \emph{voltage phase} in radian. 
Then, the following nonlinear ODE is derived in \cite{Chertkov:2011} to determine the functions $v(x)$ and $\theta(x)$: 
\begin{equation}
\left.
\begin{aligned}
\frac{d^2v}{dx^2} 
&= v\left(\frac{d\theta}{dx}\right)^2-\frac{Gp(x)+Bq(x)}{v(G^2+B^2)}
\\
-\frac{d}{dx}\left(v^2\frac{d\theta}{dx}\right) 
&= \displaystyle \frac{Bp(x)-Gq(x)}{G^2+B^2}
\end{aligned}
\right\}.
\label{eqn:ode}
\end{equation}
The constant parameters $G$ and $B$ stand for the per-unit-length conductance and susceptance [S/km].
Also, the function $p(x)$ (or $q(x)$) is the active (or reactive) power flowing into the feeder (note that $p(x)>0$ indicates the positive active power flowing to the feeder at $x$).
We will call $p(x)$ and $q(x)$ the \emph{power density functions} whose are in [W/km] and [Var/km], respectively. 
Also, as the important ancillary function in this paper, the \emph{voltage gradient} [V/km] is defined as
\begin{equation}
w(x):=\frac{d}{dx}v(x).
\label{eqn:w(x)}
\end{equation}

The ODE poses a nonlinear boundary-value problem and thus can not be analytically solved. 
The authors of \cite{Mizuta:2017,Mizuta:2018} propose an approximate solution of the problem. 
For this approximation, consider again the simple feeder model in Figure~\ref{fig:feeder1}, where $N$ number of charing stations and loads are located at $x=\xi_i\in(0,L)$ ($i=1,\ldots,N$) satisfying $\xi_{i+1}<\xi_{i}$. 
It is assumed that the in-vehicle batteries and loads are operated under unity-power factor.  
The assumption can be relaxed as in \cite{Mizuta:2018}. 
Thus, by denoting as $P_i$ the active-power discharged ($P_i>0$) (or charged (consumed) ($P_i<0$)) at $x=\xi_i$, the power density functions $p(x)$ and $q(x)$ are represented as
\begin{equation}
p(x)=\sum^{N}_{i=1}P_i\delta(x-\xi_i), \qquad q(x)=0,
\label{eqn:pq}
\end{equation}
where $\delta(x-\xi_i)$ is the Dirac's delta function supported at $x=\xi_i$.  
Then, the following approximation of solution for $w(x)$ is derived in \cite{Mizuta:2017, Mizuta:2018}:
\begin{equation}
w(x) \sim \sum_{j\in\mathcal{I}_x}P_j\frac{G}{Y^2},
\label{eqn:ssk1}
\end{equation}
where $Y:=\sqrt{G^2+B^2}$, $\mathcal{I}_x\subseteq\{1,2,\ldots,N\}$ is the set of all indexes $i$ of the locations of charging stations and loads satisfying $x<\xi_i$, $i\in\{1,\ldots,N\}$, $\xi_{N+1}:=0$. 
This implies that the voltage gradient $w(x)$ (and hence the voltage amplitude $v(x)$ through \eqref{eqn:w(x)}) can be controlled with the regulation of charging/discharging power $P_j$ by EVs. 
Based on the observation, we will introduce a systematic method of determining the values of charging and discharging power of in-vehicle batteries.

%%%%% 
\subsection{Computation of Upper Bound for Charging/Discharging Power}
\label{sec:main}

This section is a review of \cite{Yumiki} and summarizes the computation of upper bound for the synthesis of charging/discharging patterns. 
For this, let us assume that all charging stations consume non-negative power (operated at the charging mode), i.e., $P_i:=P_{\mathrm{EVs},i}\leq 0$ for all stations $i\in\{1,\ldots,N_\mathrm{sta}\}$, where $N_\mathrm{sta}$ is their total number. 
The associated location of the $i$-th station is denoted as $\xi_{\mathrm{sta},i}\in\{\xi_i,\ldots,\xi_N\}$. 
Then, from \eqref{eqn:ssk1}, the deviation of voltage amplitude at the end of the feeder ($x=L$) due to the charging of EVs  is approximately estimated in \cite{Yumiki} as
\begin{equation}
dv(L) = \sum_{i=1}^{N\sub{sta}} \frac{G}{Y^2}|P_{\mathrm{EVs},i}|\xi_{\mathrm{sta},i}.
\label{eqn:dv}
\end{equation}
See \cite{Yumiki} for its derivation. 
This shows that $dv(L)$ is determined with the amounts of charging power and locations of the stations. 
The evaluation formula \eqref{eqn:dv} can work for the case, where all the stations are operated at the discharging mode (i.e., $P_{\mathrm{EVs},i}\geq 0$ for all $i$). 

We now describe the method for computing an upper bound of the synthesis of charging (or discharging) patterns \cite{Yumiki}. 
The method is based on the simple evaluation of \eqref{eqn:dv} and effective for regulating $dv(L)$ at its pre-defined acceptable value, which we denote by $dV_\mathrm{limit}$. 
The method eventually works if $dv(L)>dV_\mathrm{limit}$. 
For this, we use a common parameter for each station, denoted by $\alpha\in[0,1]$, and refine \eqref{eqn:dv} as
\begin{align}
dv(L)_\alpha &:= \sum_{i=1}^{N\sub{sta}}\frac{G}{Y^2}|\alpha P_{\mathrm{EVs},i}^\mathrm{max}|\xi_{\mathrm{sta},i}
= \alpha\times dv(L).
\label{eqn:dv_a}
\end{align}
This $dv(L)_\alpha$ represents a modification of the measure of \eqref{eqn:dv} by uniformly regulating (decreasing) the maximum power of every station.  
Then, by taking $dv(L)_\alpha = dV_\mathrm{limit}$ as the physical specification of voltage, $\alpha$ is determined as
\begin{equation}
\alpha=\frac{dV_\mathrm{limit}}{dv(L)}.
\label{eqn:alpha}
\end{equation}
For computing the upper bound, $\alpha_\mathrm{cha}$ (or $\alpha_\mathrm{discha}$) is calculated with \eqref{eqn:alpha} and the pre-defined value $dV_\mathrm{cha, limit}$ at the charging mode (or $dV_\mathrm{discha, limit}$ at the discharging mode). 
Hence, it is possible to compute the upper bound as $\{-{\alpha_\mathrm{cha}} P_{\mathrm{EVs},i}^\mathrm{max} : i=1,\ldots,N_\mathrm{sta}\}$ (or $\{{\alpha_\mathrm{discha}} P_{\mathrm{EVs},i}^\mathrm{max} : i=1,\ldots,N_\mathrm{sta}\}$) so that the deviation of voltage amplitude at the end point of the feeder due to charging (or discharging) of EVs can be bounded at $dV_\mathrm{cha, limit}$ (or $dV_\mathrm{discha, limit}$).

Here, it should be emphasized that the computation of upper bound becomes feasible when the cooperative management works for DSO and EV-sharing operator. 
It requires both information of distribution grid and mobility system. 
Like Figure~\ref{fig:system}, from EV-sharing operator, DSO receives prediction data of the number of EVs at charging stations in a distribution feeder. 
By using \eqref{eqn:dv}, DSO judges whether the deviation of voltage amplitude at the end point of the feeder is smaller than $dV\sub{limit}$ or not. 
For this, it is firstly supposed that all the number of in-vehicle batteries connected to every station can charge (or discharge) with the maximum power $-P_{\mathrm{EVs},i}^\mathrm{max}\,(<0)$ and $P_{\mathrm{EVs},i}^\mathrm{max}\,(>0)$. 
If $dv(L)\leq dV\sub{limit}$, then the in-vehicle batteries under the sharing service can charge or discharge with their maximum power at every station. 
Otherwise, namely, $dv(L)> dV\sub{limit}$, then DSO computes the upper bound as shown above and sends it to EV-sharing operator in order to keep the voltage level.

%%%%
%%%%
\section{Proposed Autonomous Vehicle-to-Grid Design}
\label{sec:autonomou}

This section aims to describe the main idea of this paper: to propose the novel control system referred to as autonomous V2G design for providing the multi-objective AS: PFC reserve and voltage regulation.   
In this design, charging or discharging power by EVs distributed for multiple stations is regulated with the \emph{local}  measurement of frequency, while the maximum amount of charging or discharging power is determined \emph{globally} using the method in Section~\ref{sec:main}. 
In this sense, the distributed EVs are cooperative with the voltage regulation of a distribution grid. 

\begin{figure}[t]
\centering
\includegraphics[width=.6\textwidth]{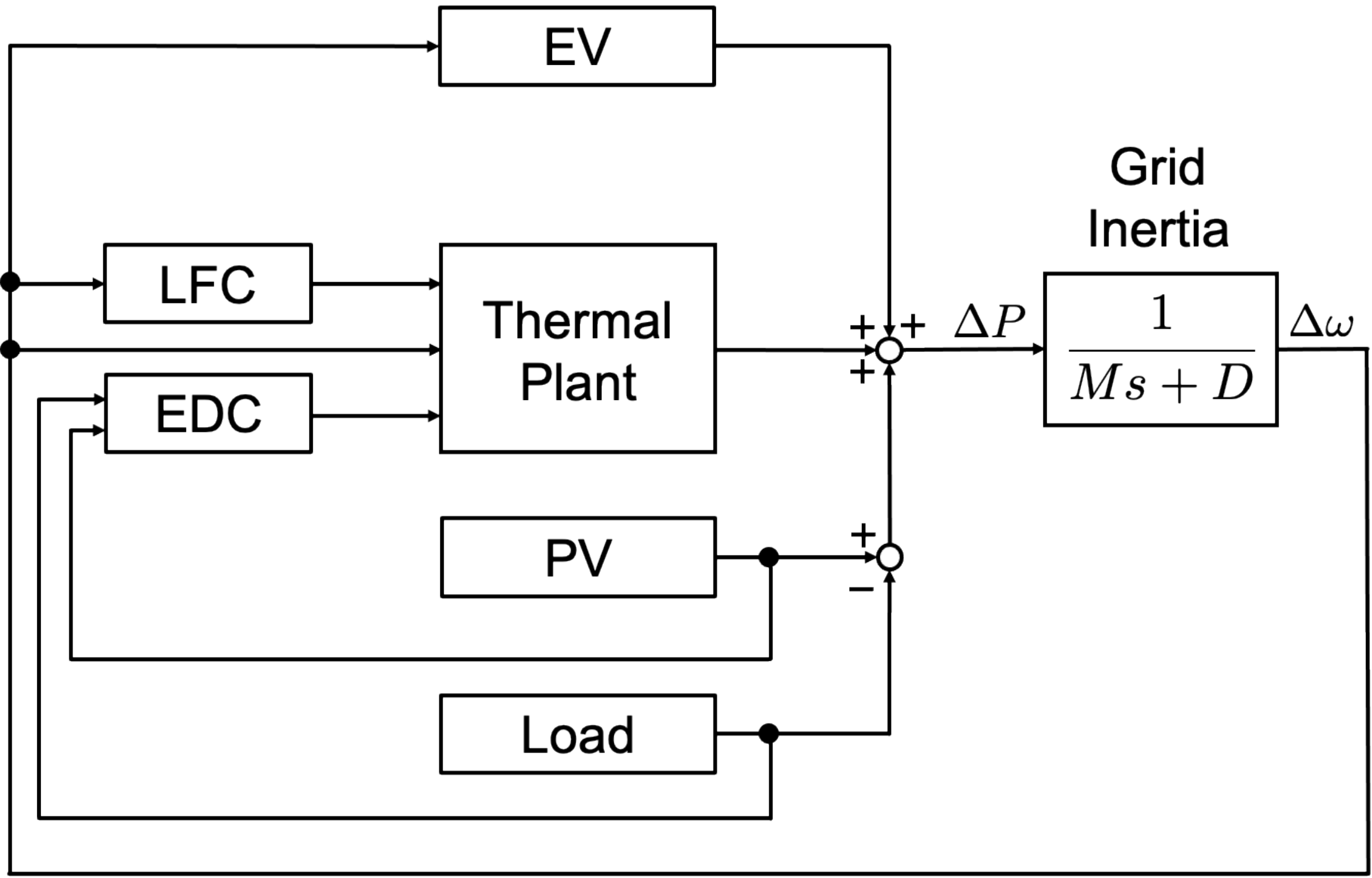} %eps}
\caption{%
Block diagram for frequency dynamics of a transmission grid 
}%
\label{fig:f}
\end{figure}

For the multi-objective AS, in the rest of this paper we consider the conventional hierarchical structure of transmission and distribution grids: in the high-level transmission grid or the associated TSO, the frequency deviation is determined in a dynamic manner: and in a low-level distribution feeder or DSO, the voltage profile is determined in a static manner. 
For the distribution feeder, we utilize the ODE model in order to assess and regulate the voltage profile. 
For the transmission grid, we introduce the block diagram for frequency dynamics in Figure~\ref{fig:f} that is based on \cite{Toda:2017,Mizuta:2019}. 
The deviation $\Delta \omega$ of the grid's angular frequency from the nominal, $50\,\U{Hz}\times 2\pi$ in this paper, is determined with the net imbalance $\Delta P$ of supply and demand in the entire grid, the inertia constant $M$ of the grid, and its damping coefficient $D$. 
The net imbalance $\Delta P$ is calculated with the output of thermal plant determined by economic dispatch control (EDC) \cite{Andersson}, load frequency control (LFC) \cite{Andersson}, load in the hierarchical grids, power generation by photo-voltaic (PV) generation units, and charging/discharging power of EVs. 
The model of thermal power plant with turbine and governor is also based on \cite{Toda:2017,Mizuta:2019}. 

\begin{figure}[t]
\centering
\includegraphics[width=.5\textwidth]{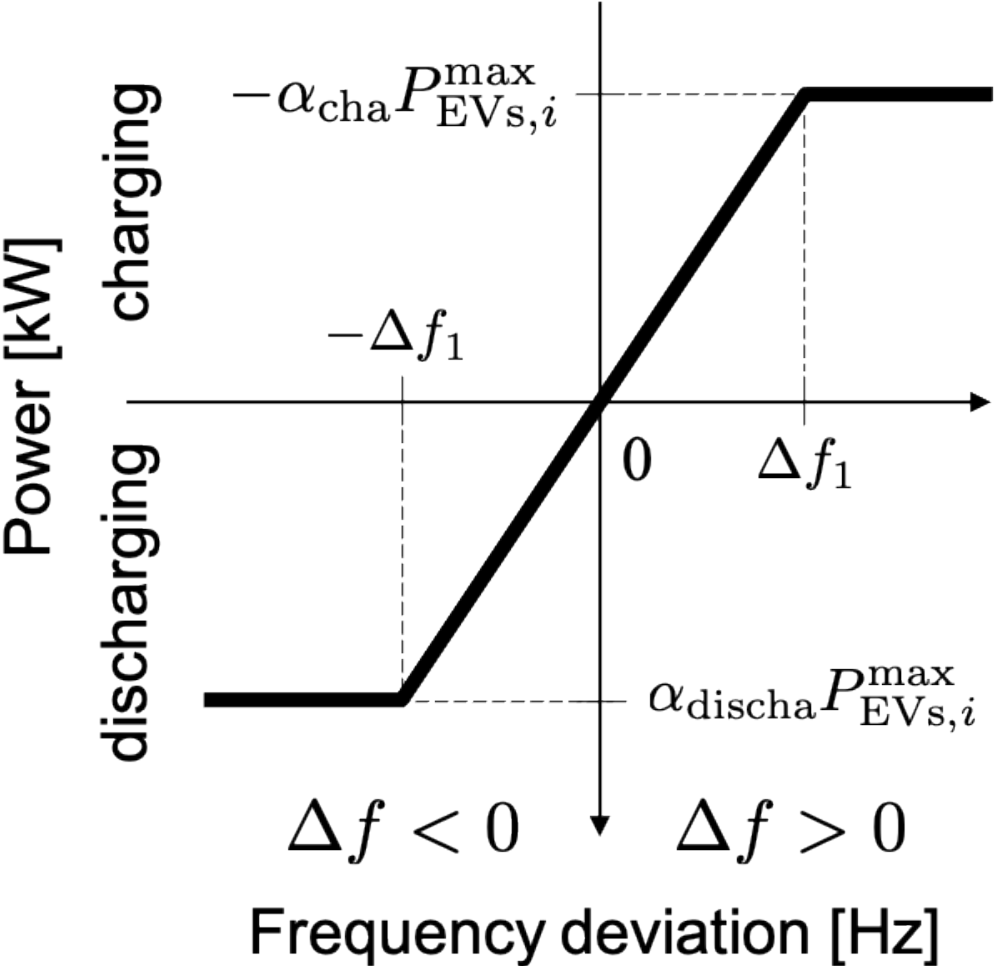} %eps}
\caption{%
Charging/discharging regulation with the droop-type characteristics against the frequency deviation for autonomous vehicle-to-grid design
}%
\label{fig:droop}
\end{figure}

Now, we are in position to describe the autonomous V2G design for determining active power output by EVs. 
Specifically, we adopt from \cite{Ota:2012} the autonomous scheme based on distributed EVs for realizing the fast responsiveness to TSO. 
Here, it is supposed that the target distribution feeder has $N\sub{sta}$ charging stations labeled by integer $i$ as introduced in Section~\ref{sec:main}. 
Then, DSO computes the upper bound of charging (or discharging) power as $-\alpha_\mathrm{cha} P_{\mathrm{EVs},i}^\mathrm{max}$ (or $\alpha_\mathrm{discha} P_{\mathrm{EVs},i}^\mathrm{max})$ at $i$-th station connected to the feeder at $x=\xi_{\mathrm{sta},i}$. 
In this paper, we propose that the charging/discharging power $P_{\mathrm{EVs},i}$ at $i$-th station connected to the feeder at $x=\xi_{\mathrm{sta},i}$ is regulated with the droop-type characteristics as in \cite{Ota:2012}. 
To do this, the frequency deviation $\Delta f:=\Delta\omega/(2\pi)$ is locally measured at the connection terminal, and $P_{\mathrm{EVs},i}$ is then determined as shown in Figure~\ref{fig:droop},
\begin{equation}
\makebox[-.5em]{}P_{\mathrm{EVs},i}=
\left\{
\begin{array}{rll}
{-\alpha_\mathrm{cha} P_{\mathrm{EVs}, i}^{\max }}, 
& \textrm{if} & \Delta f_{1}\leq \Delta f
\\\noalign{\vskip 2mm}
{K_\mathrm{cha} \Delta f}, 
& \textrm{else if} & 0\leq \Delta f < \Delta f_{1}
\\\noalign{\vskip 2mm}
{K_\mathrm{discha} \Delta f}, 
& \textrm{else if} & -\Delta f_{1} \leq \Delta f<0
\\\noalign{\vskip 2mm}
{\alpha_\mathrm{discha} P_{\mathrm{EVs}, i}^{\mathrm{max}}}, 
& \textrm{else},
\end{array}
\right.
\label{eqn:droop2}
\end{equation}
where $K_\mathrm{cha}$ (or $K_\mathrm{discha}$) is calculated with 
${\alpha_\mathrm{cha} P_{\mathrm{EVs}, i}^{\max }}$ (or ${\alpha_\mathrm{discha} P_{\mathrm{EVs}, i}^{\max }}$) and the new parameter $\Delta f_{1}$. 
The regulation scheme \eqref{eqn:droop2} indicates that EVs connected to $i$-th station can charge and discharge within the pre-defined range of frequency, $[50\,\U{Hz}-\Delta f_{1}$, $50\,\U{Hz}+\Delta f_{1}]$, and thus they behave in a similar manner as the thermal power plant with PFC \cite{Andersson}. 
The scheme is based on the local measurement of the grid's frequency, while the maximum amount of the active power output is managed globally using the upper bound in Section~\ref{sec:main}. 
Thus, the EVs distributed for multiple stations are capable of providing the PFC reserve, while they are cooperative with the voltage regulation.  

Here, we discuss the capability of providing the PFC reserve to TSO. 
The capability measure in the distribution feeder at the charging (or discharging) mode, denoted as $\Delta P_\mathrm{cha/Hz}$ (or $\Delta P_\mathrm{discha/Hz}$), is defined in this paper as follows:
\begin{equation}
\left.
\begin{aligned}
\Delta P_\mathrm{cha/Hz}
&:= -{\frac{\alpha_\mathrm{cha}}{\Delta f_{1}}} {\displaystyle \sum_{i=1}^{N\sub{sta}} P_{\mathrm{EVs},i}^\mathrm{max}}
\\
\Delta P_\mathrm{discha/Hz}
&:=
{\frac{\alpha_\mathrm{discha}}{\Delta f_{1}}} {\displaystyle \sum_{i=1}^{N\sub{sta}} P_{\mathrm{EVs},i}^\mathrm{max}}
\end{aligned}
\right\}.
\label{eqn:reserve}
\end{equation}
Under a fixed value of $\alpha_\mathrm{cha}$ (or $\alpha_\mathrm{discha}$), the capability measure $\Delta P_\mathrm{cha/Hz}$ (or $\Delta P_\mathrm{discha/Hz}$) is inversely proportional to the parameter $\Delta f_{1}$. 
The measures \eqref{eqn:reserve} indicate the regulation reserve of active power by EVs in the distribution feeder per the unit frequency deviation and are closely related to the frequency characteristic requirement in \cite{Rebours} as a parameter of PFC, to which is referred as the frequency characteristic of a control area. 

It is valuable to compare the schemes proposed by \cite{Ota:2012} and in \eqref{eqn:droop2}. 
In \cite{Ota:2012}, distributed EVs at $i$-th station can charge (or discharge) up to the maximum determined by EV-sharing operator, i.e., $-P_{\mathrm{EVs},i}^\mathrm{max}$ (or $P_{\mathrm{EVs},i}^\mathrm{max}$). 
In the proposed method, distributed EVs at $i$-th station can charge (or discharge) up to the upper bound determined cooperatively by EV-sharing operator and DSO, i.e., $-{\alpha_\mathrm{cha} P_{\mathrm{EVs}, i}^{\max }}$ (or ${\alpha_\mathrm{discha} P_{\mathrm{EVs}, i}^{\max }}$). 
Because of $\alpha_\mathrm{cha}$ and $\alpha_\mathrm{discha}$ less than or equal to unity, the capability measure of providing the PFC reserve for \cite{Ota:2012} is larger than those in \eqref{eqn:droop2}. 
However, as described in Section~\ref{sec:main}, the proposed scheme provides a cooperative operation of distributed EVs for the voltage regulation. 
No measure for the multi-objective AS is reported to the best of our survey, and its unified definition remains to be solved is in our future research. 
This is related to the current interest of the cooperation of DSO and TSO: see, e.g., \cite{Martini}. 

As the end of this section, we summarize the autonomous V2G design for providing the multi-objective AS, consisting of the following three steps {\bf T1} to {\bf T3}:
\begin{itemize}
\item[\bf T1] DSO updates the prediction data on the number of EVs at charging stations 
from EV-sharing operator every pre-defined period. 
One example of the period is 15\,minutes in \cite{Masegi:2018} where EV-sharing operator receives the information on reservations by users every 15\,minutes.  
\item[\bf T2] By using the operational data on EVs and the method described in Section~\ref{sec:main}, DSO computes the upper bound of charging/discharging of EVs at each station in terms of the voltage management. 
In this paper, we assume that DSO determines the value of $dV_\mathrm{cha, limit}$ (or $dV_\mathrm{discha, limit}$) and thereby the upper bound for the synthesis of charging (or discharging) patterns as $-{\alpha_\mathrm{cha} P_{\mathrm{EVs}, i}^{\max }}\,(<0)$ (${\alpha_\mathrm{discha} P_{\mathrm{EVs}, i}^{\max }}\,(>0)$). 
\item[\bf T3] EVs at each station can charge and discharge within the upper bound by DSO.  
This operation is conducted only with the local measurement of the grid's frequency. 
No exchange of information between any charging stations is required. 
\end{itemize}

%%%%
%%%%
\section{Effectiveness Evaluation}
\label{sec:exp}

The aim of this section is to evaluate the effectiveness of the proposed design using full-digital simulations of both the frequency dynamics and the distribution voltage profile. 

%%%
\subsection{Simulation Setting}
\label{sec:Simulation Setting}

\begin{figure}[t]
\centering
\includegraphics[width=.7\textwidth]{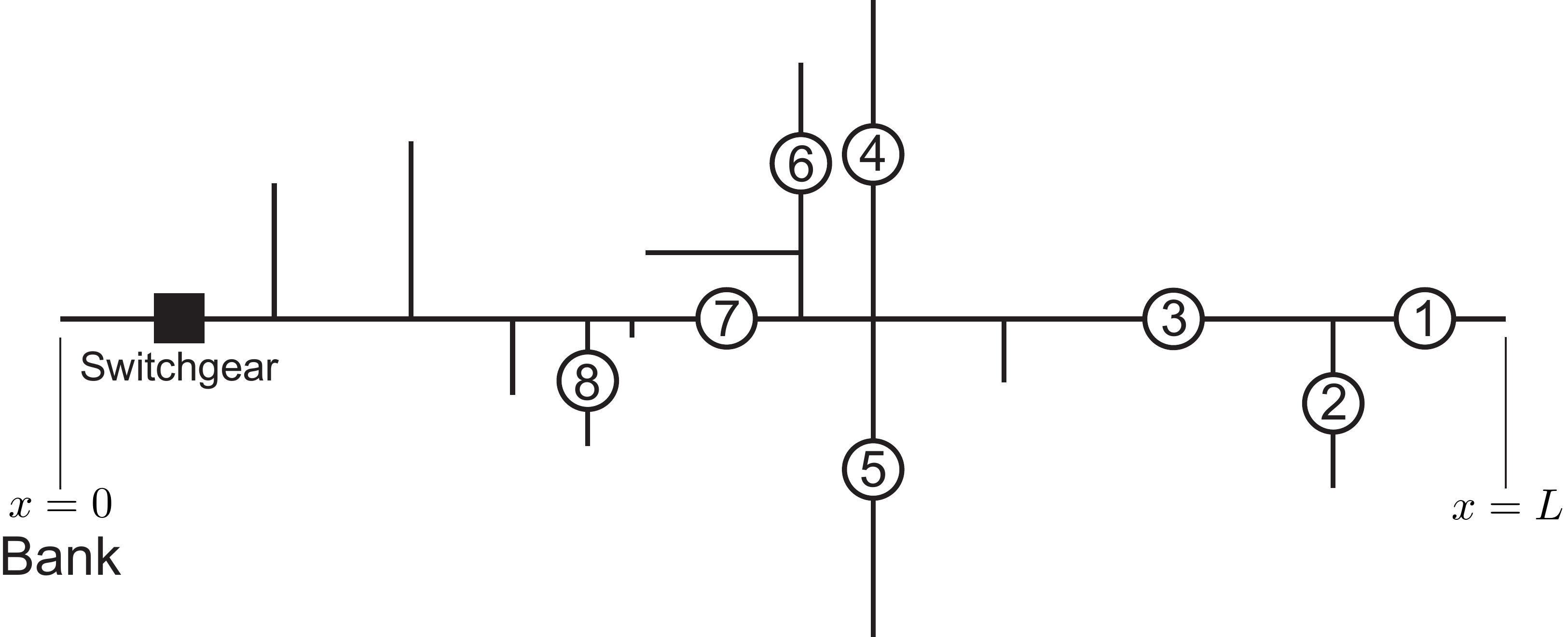}
\caption{%
Model of multiple feeders based on a practical distribution grid of residential area in Japan.
The 8 charging stations are installed and denoted by the circled number: see \cite{Mizuta:2018} for details.
}%
\label{fig:model2}
\includegraphics[width=.6\textwidth]{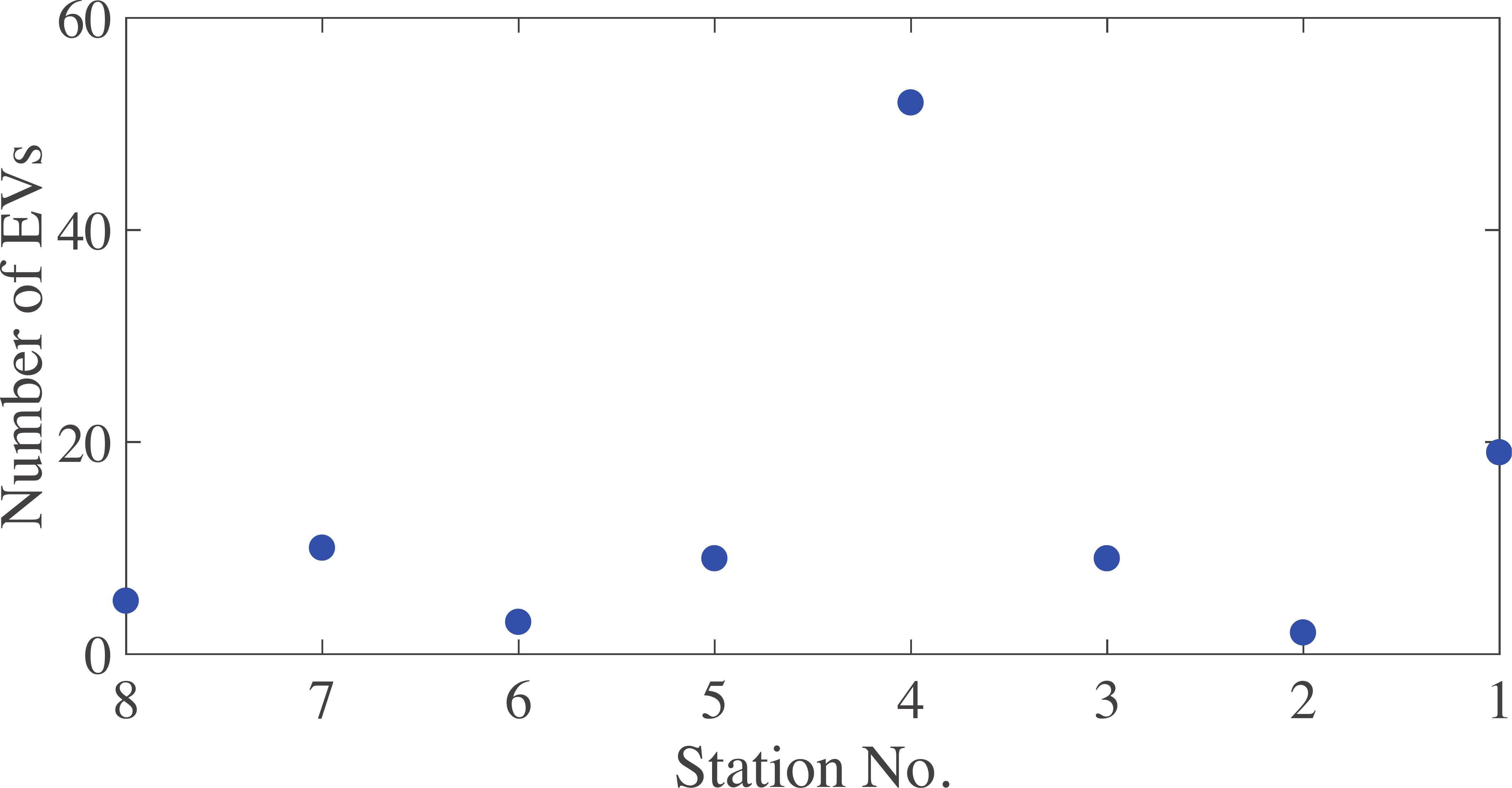} %eps}
\caption{%
Number of electric vehicles (EVs) at the 8 charging stations in the distribution feeder of Figure~\ref{fig:model2}.
}%
\label{fig:EV}
\end{figure}

First, we describe the simulation setting of power grids and mobility system.
The setting of transmission grid in Figure~\ref{fig:f} is based on \cite{Mizuta:2019}, where we assume that the transmission grid is spanned in a geographical region with a population of approximately 9 million customers, and that the grid's capacity is about $8.3\,\U{GVA}$. 
It is also assumed that the PV generation is installed as a mega solar plant with 20\% against the grid's capacity. 
Here, for simplicity of the analysis, the smoothing effect of introduction of PV is not considered. 
The PV generation as well as the load consumption changes in an aperiodic manner as time goes on. 
These time-varying inputs cause the frequency dynamics of the grid, and no event or test case is considered in the following simulations. 
The nominal frequency of the grid is $50\,\U{Hz}$, the inertia constant and damping coefficient in time are $9\,\U{s}$ and $2\,\U{s}$, and the other parameters of the plant model are the same as in \cite{Mizuta:2019}.  
As the distribution grid shown in Figure~\ref{fig:model2}, the model of multiple feeders is used in \cite{Mizuta:2018} and based on a practical distribution grid of residential area in western Japan. 
We assume that the secondary voltage at the bank is regulated at $6.6\,\U{kV}$. 
The rated capacity of the bank is set at $12\,\U{MVA}$, and the feeder's resistance (or reactance) at $0.227\,\U{\Omega/km}$ (or $0.401\,\U{\Omega/km}$). 
See \cite{Mizuta:2018} for details. 
It is supposed that 600 same feeders of the distribution grid are connected to the transmission grid. 
For the mobility system, we use synthetic operation data on EVs in a sharing service for the numerical evaluation: see \cite{Yumiki} for details. 
The number of EVs at the 8 stations that does not change for 15\,minutes from 12 o'clock are shown in Figure~\ref{fig:EV}. 
The constant number is used in the numerical simulations over 60\,minutes under time-varying PV generation and load consumption.  

\begin{figure*}[t]
\centering
\begin{minipage}{.495\hsize}
\centering
\hspace*{-4mm}\includegraphics[width=.875\hsize]{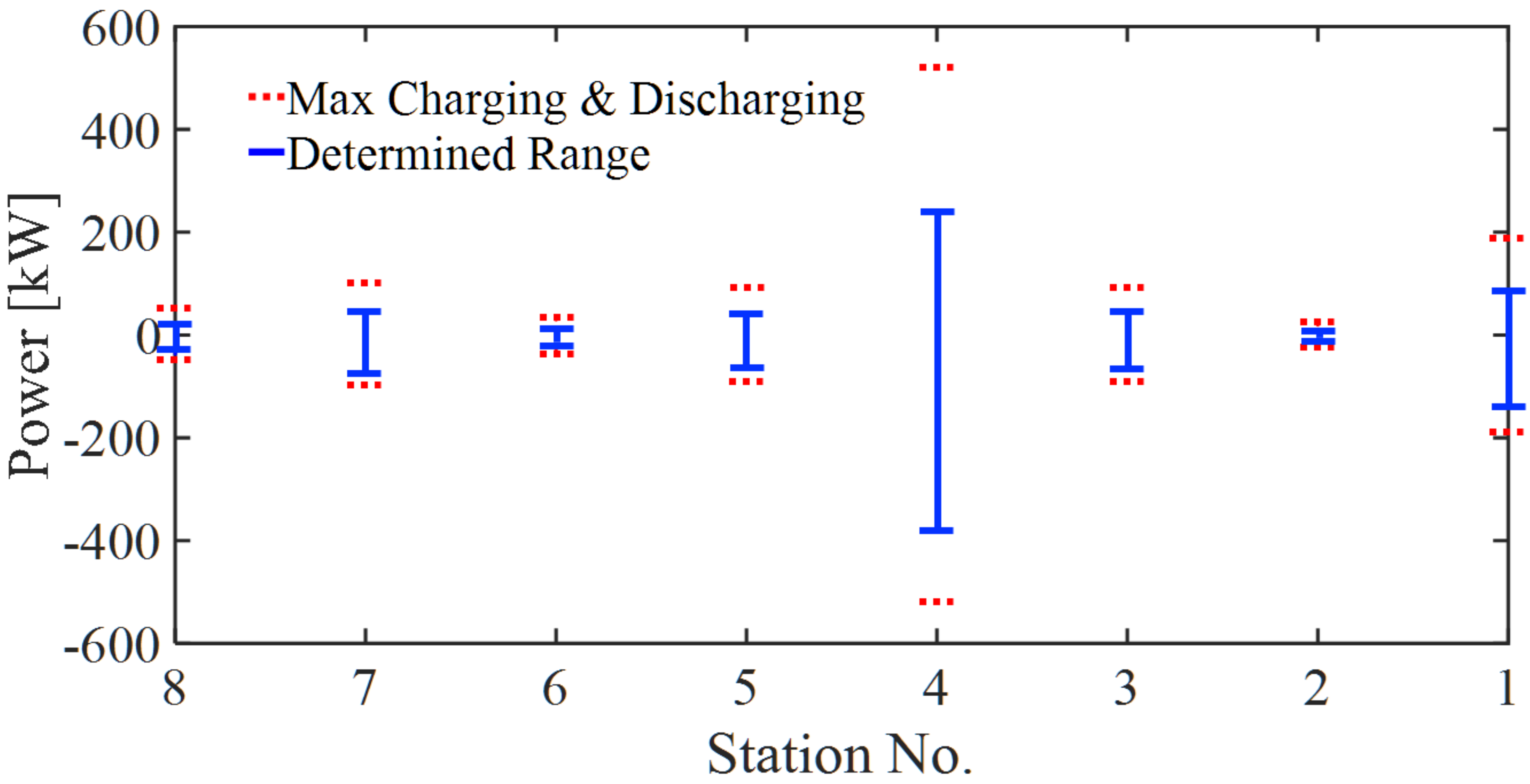} %{./figsPt1/power.eps}
\includegraphics[width=\hsize]{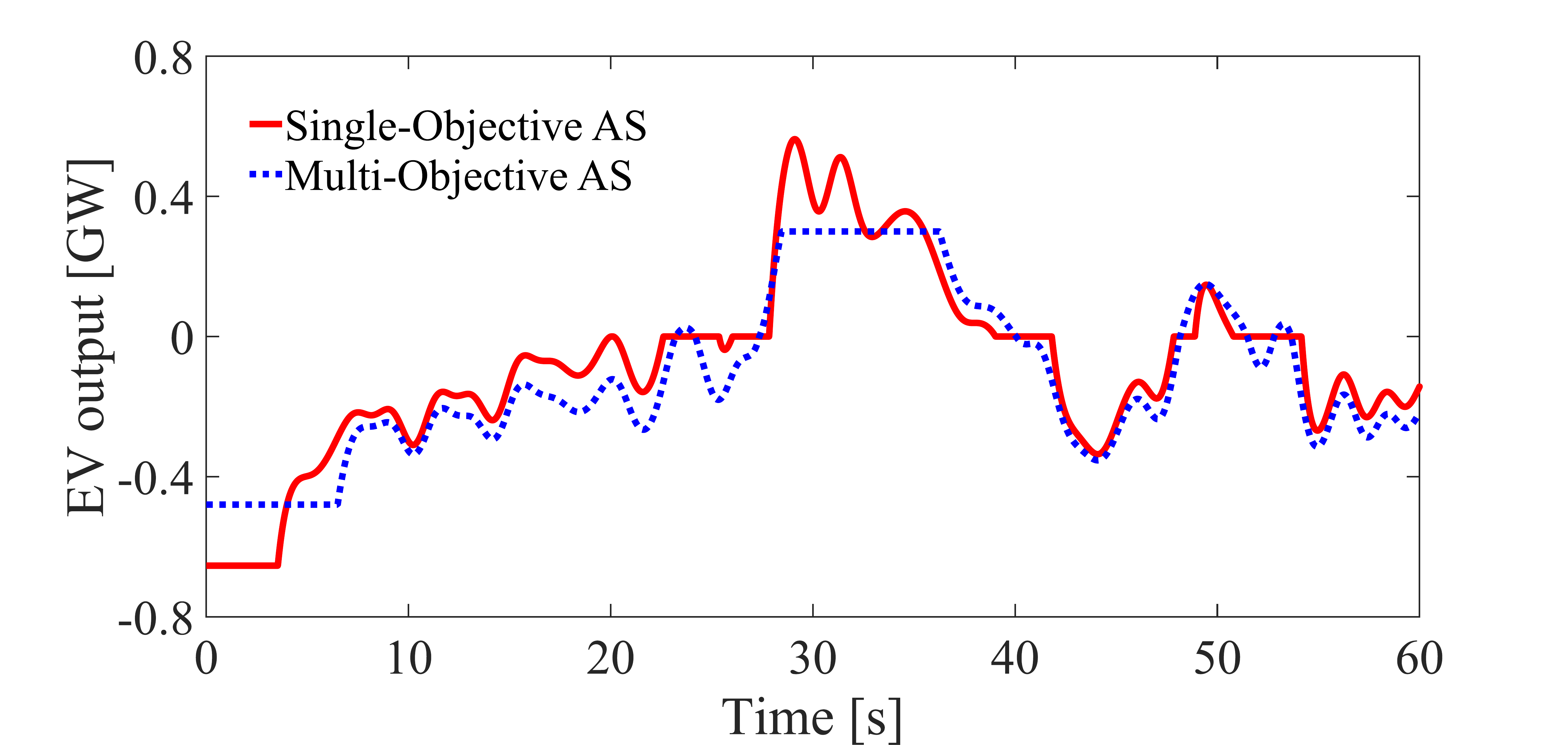} %{./figsPt1/EV1output.eps}
\includegraphics[width=\hsize]{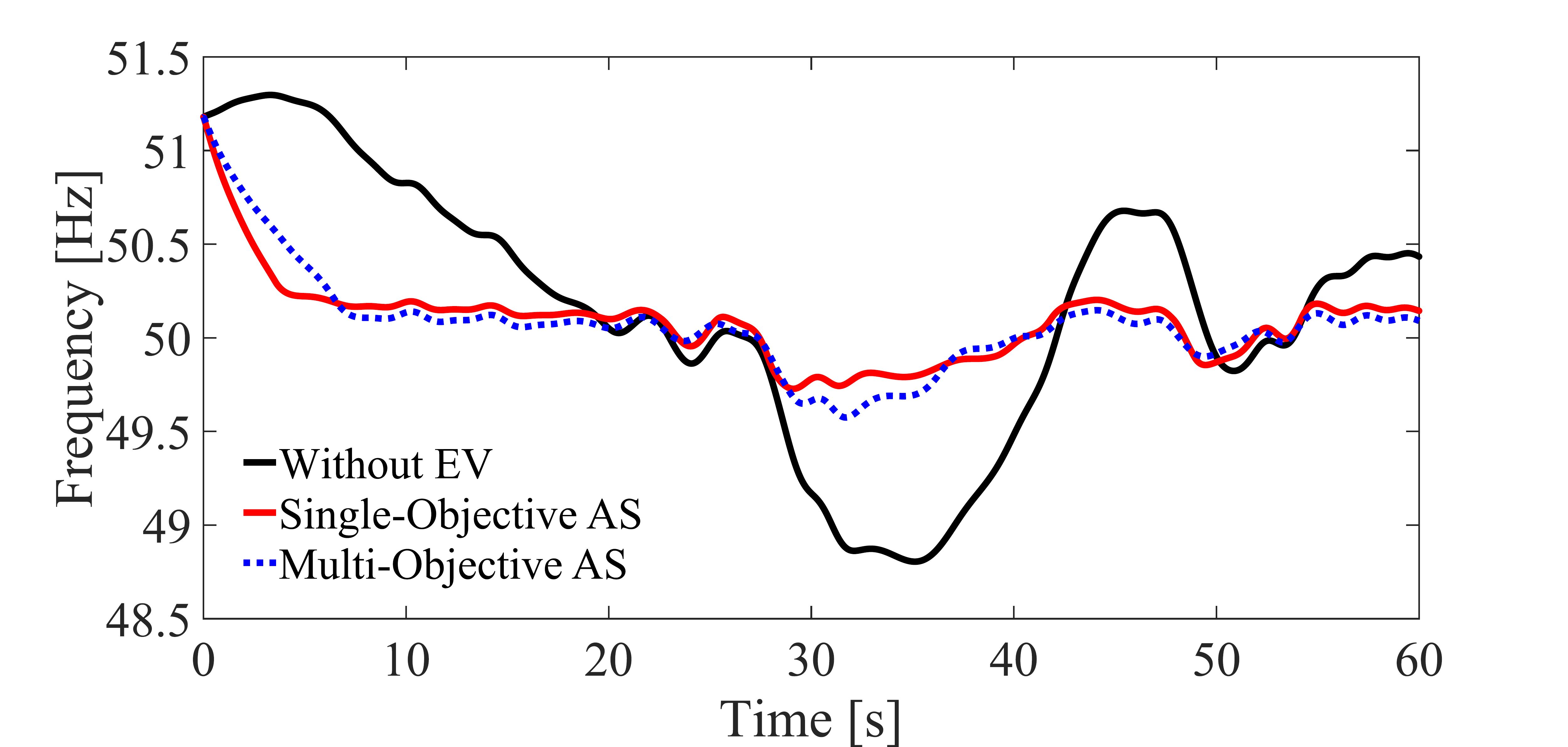} %{./figsPt1/f1.eps}
\subcaption{%
$(dV_\mathrm{cha, limit}, dV_\mathrm{discha, limit})=(80\,\U{V},\,50\,\U{V})$
}%
\label{fig:Pref=0}
\end{minipage}
\begin{minipage}{.495\hsize}
\centering
\hspace*{-4mm}\includegraphics[width=.875\hsize]{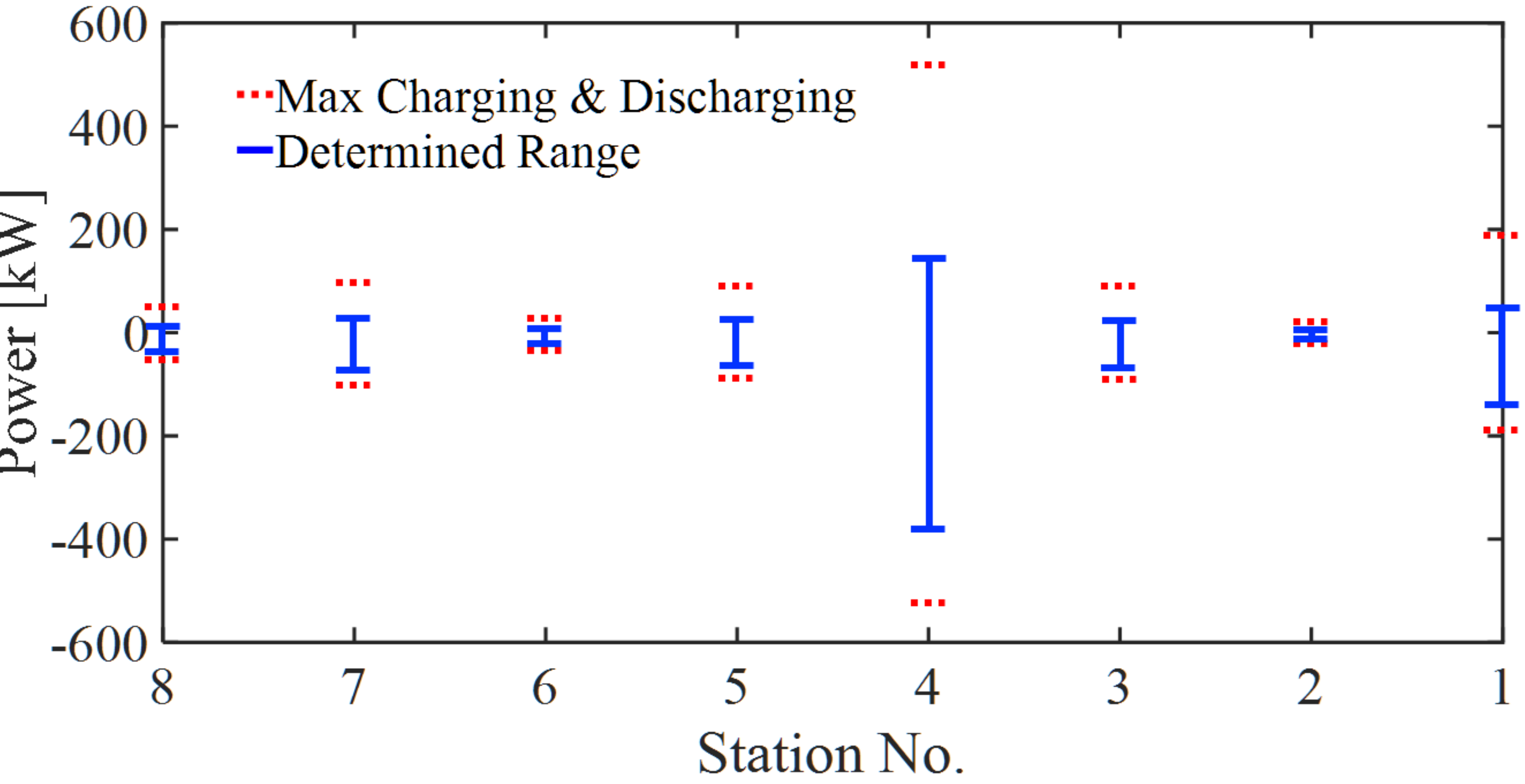} %{./figsPt1/power2.eps}
\includegraphics[width=\hsize]{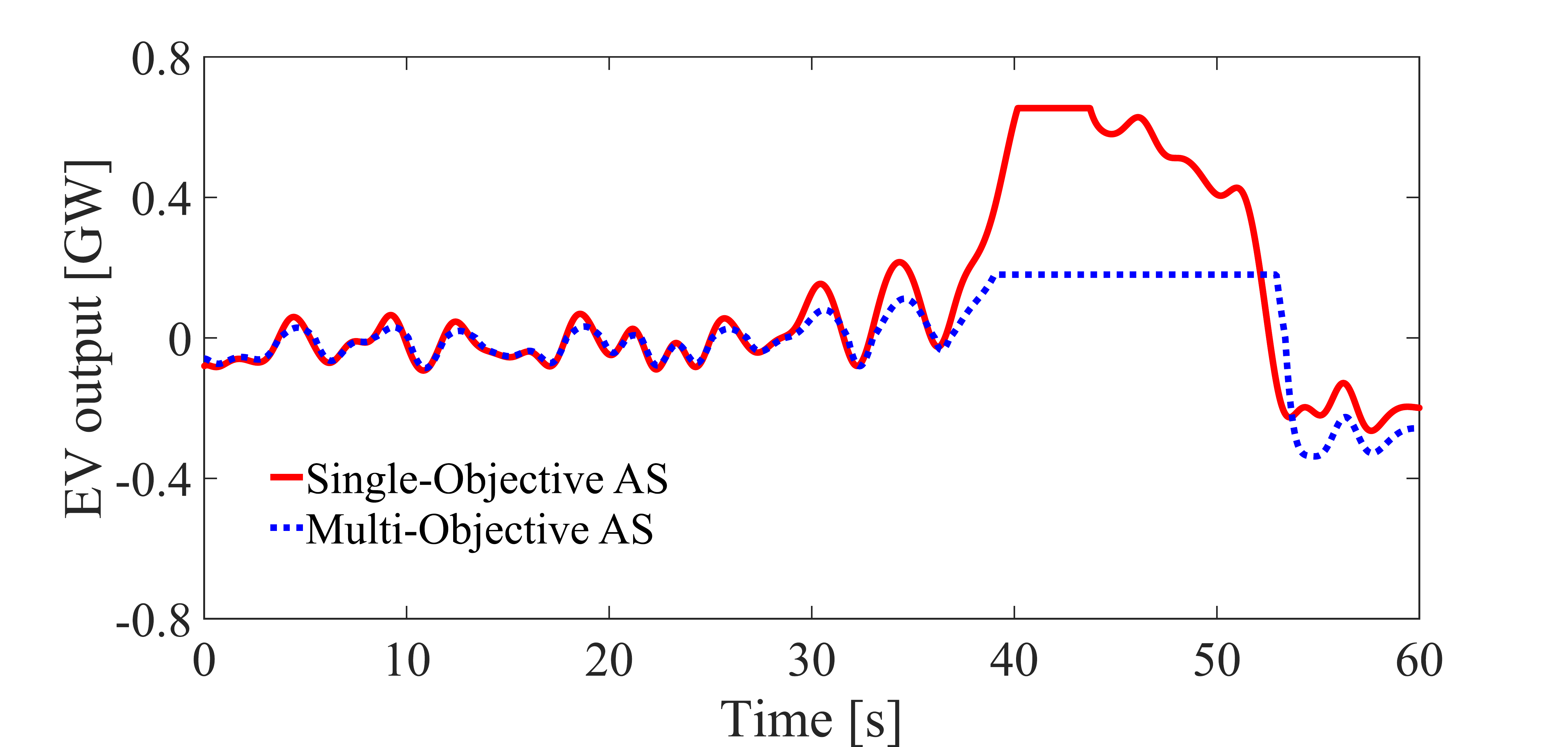} %{./figsPt1/EV3output.eps}
\includegraphics[width=\hsize]{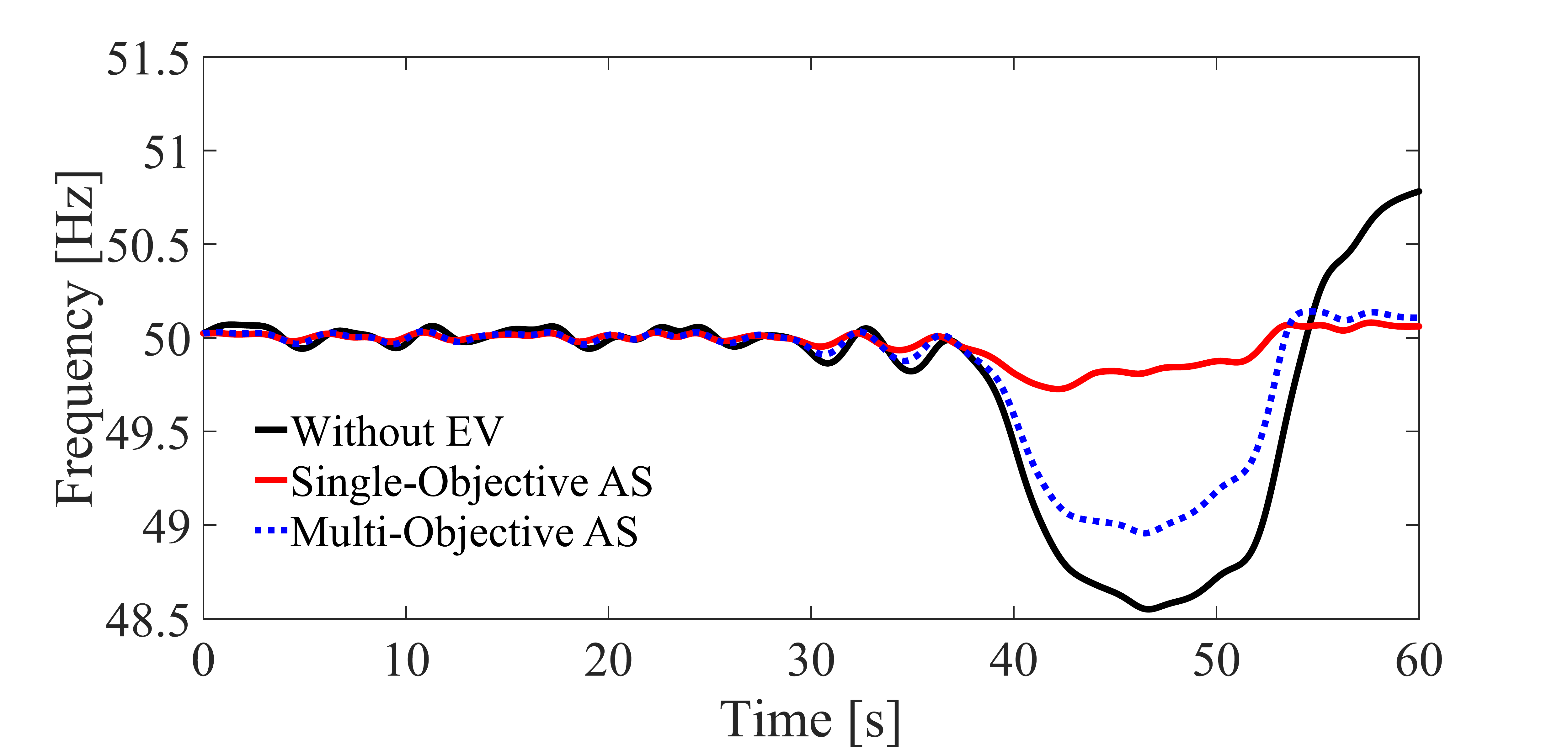} %{./figsPt1/f3.eps}
\subcaption{%
$(dV_\mathrm{cha, limit}, dV_\mathrm{discha, limit})=(80\,\U{V},\,30\,\U{V})$
}%
\label{fig:Pref=5}
\end{minipage}
\caption{%
Numerical evaluation of autonomous vehicle-to-grid design for multi-objective ancillary service--I: 
(top) synthesis results of upper bounds for charging/discharging patterns of in-vehicle batteries; 
(middle) time responses of total output power from electric vehicles (EVs); 
(bottom) time responses of the grid's frequency with/without EVs. 
}%
\label{fig:demo1}
\end{figure*}

\begin{figure*}[t]
\centering
\begin{minipage}{.495\hsize}
\centering
\includegraphics[width=\hsize]{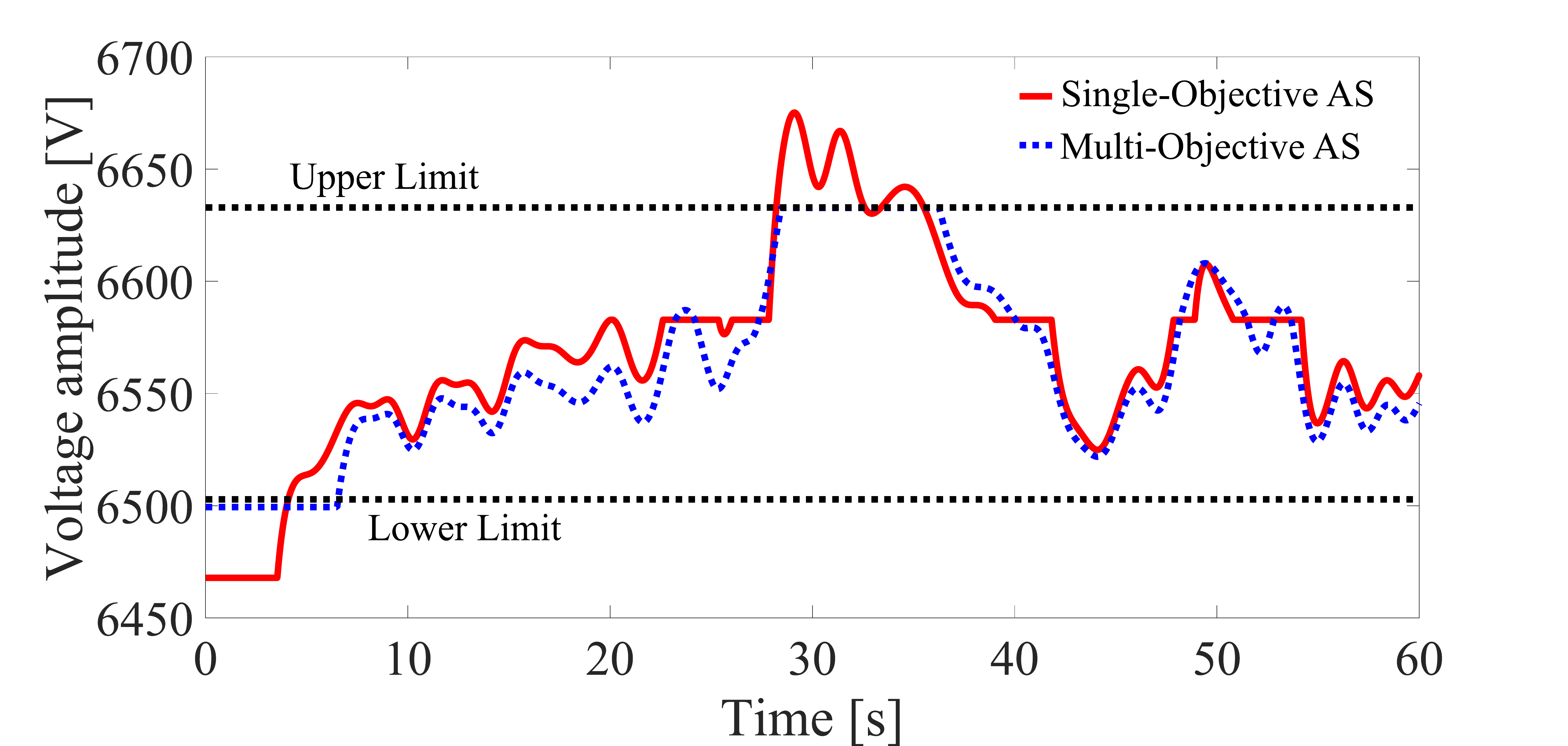} %{./figsPt1/v2L.eps}
\includegraphics[width=\hsize]{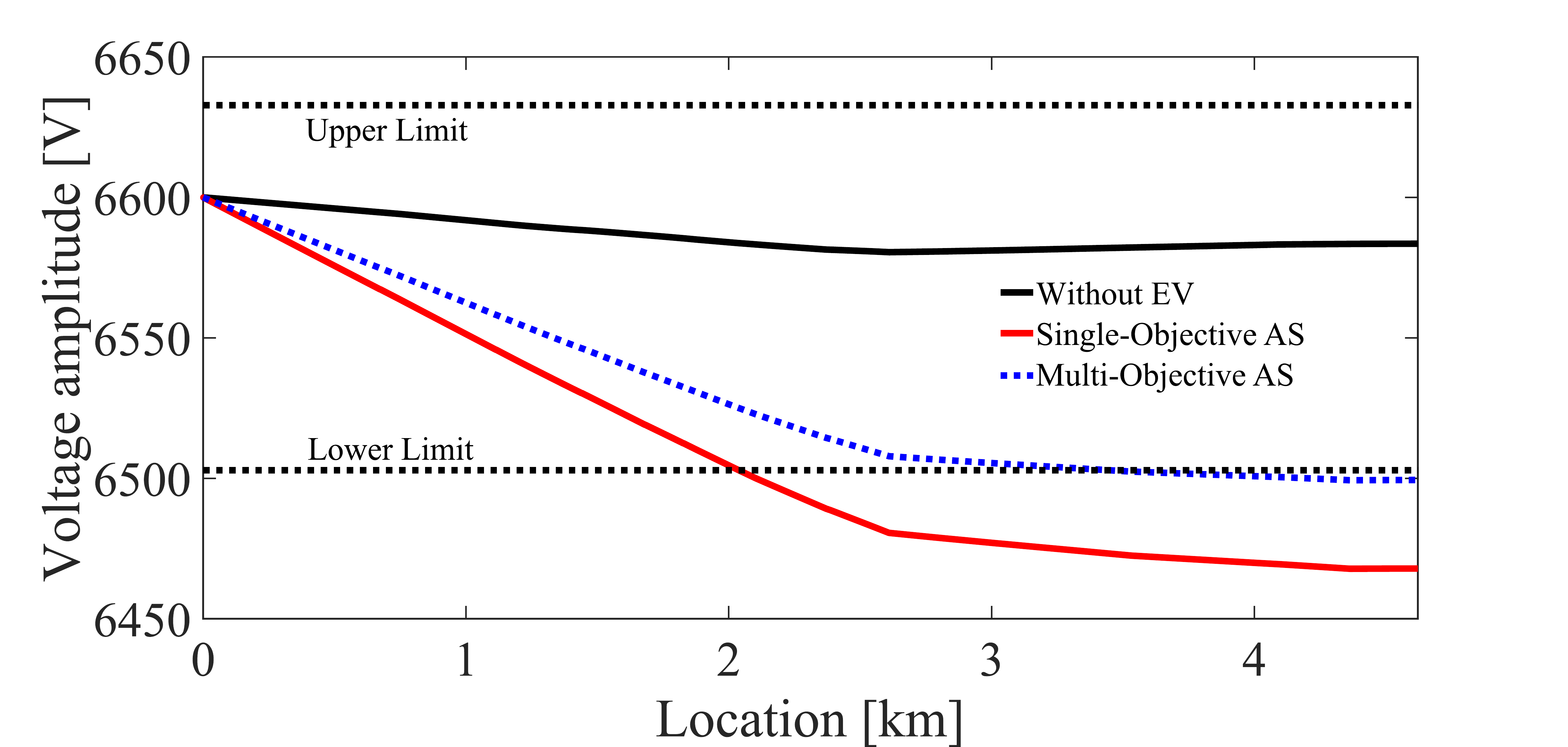} %{./figsPt1/v_2s.eps}
\subcaption{%
$(dV_\mathrm{cha, limit}, dV_\mathrm{discha, limit})=(80\,\U{V},\,50\,\U{V})$
}%
\label{fig:Pref=0}
\end{minipage}
\begin{minipage}{.495\hsize}
\centering
\includegraphics[width=\hsize]{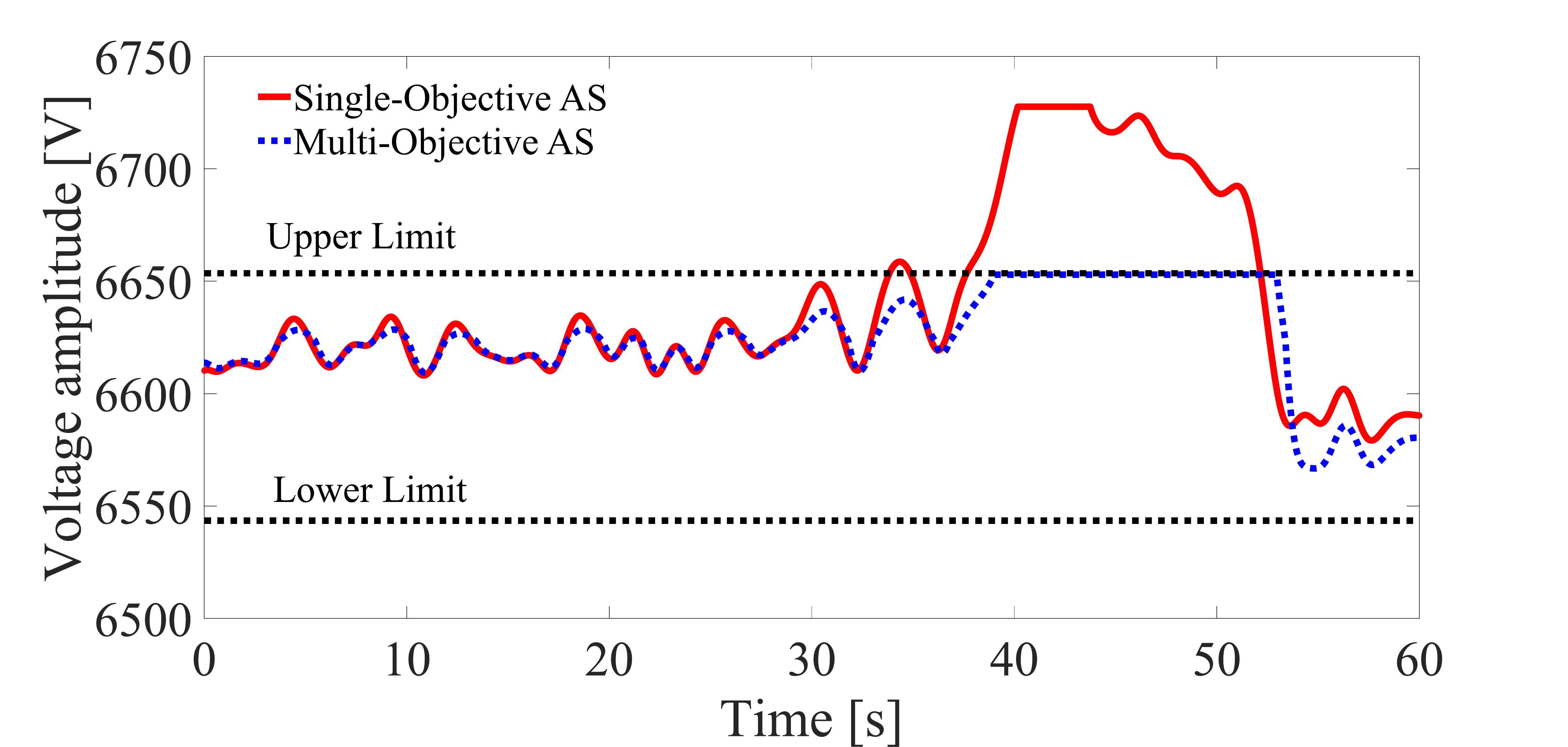} %{./figsPt1/v4L.eps}
\includegraphics[width=\hsize]{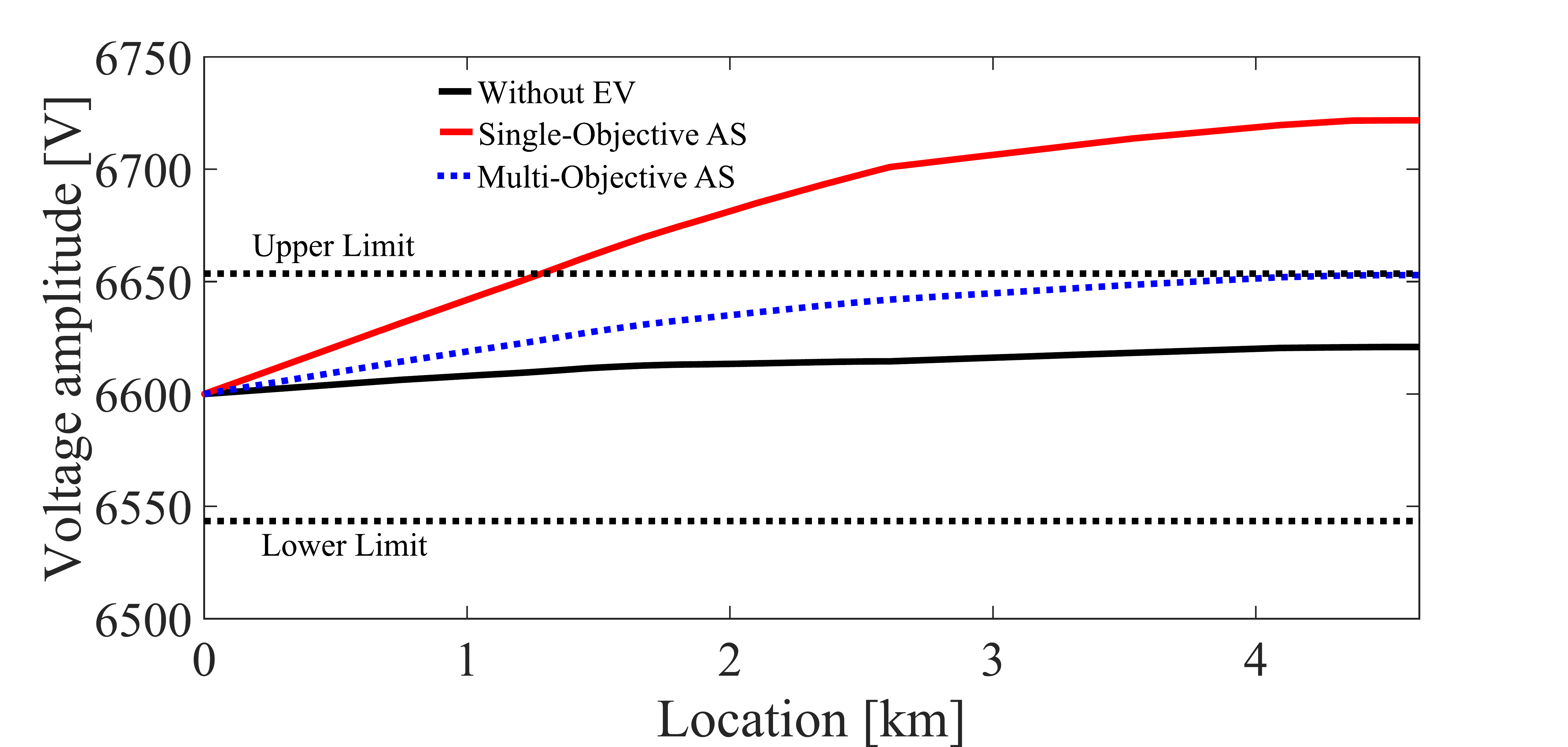} %{./figsPt1/v_40s.eps}
\subcaption{%
$(dV_\mathrm{cha, limit}, dV_\mathrm{discha, limit})=(80\,\U{V},\,30\,\U{V})$
}%
\label{fig:Pref=5}
\end{minipage}
\caption{%
Numerical evaluation of autonomous V2G design for multi-objective ancillary service--II: 
(top) time series of distribution voltage at $x=L$ and 
(bottom) spatial profile of distribution voltage at time $2\,\U{s}$ for (a) and at $40\,\U{s}$ for (b). 
}%
\label{fig:demo2}
\end{figure*}

\begin{figure*}[t]
\begin{minipage}{.495\hsize}
\centering
\includegraphics[width=\hsize]{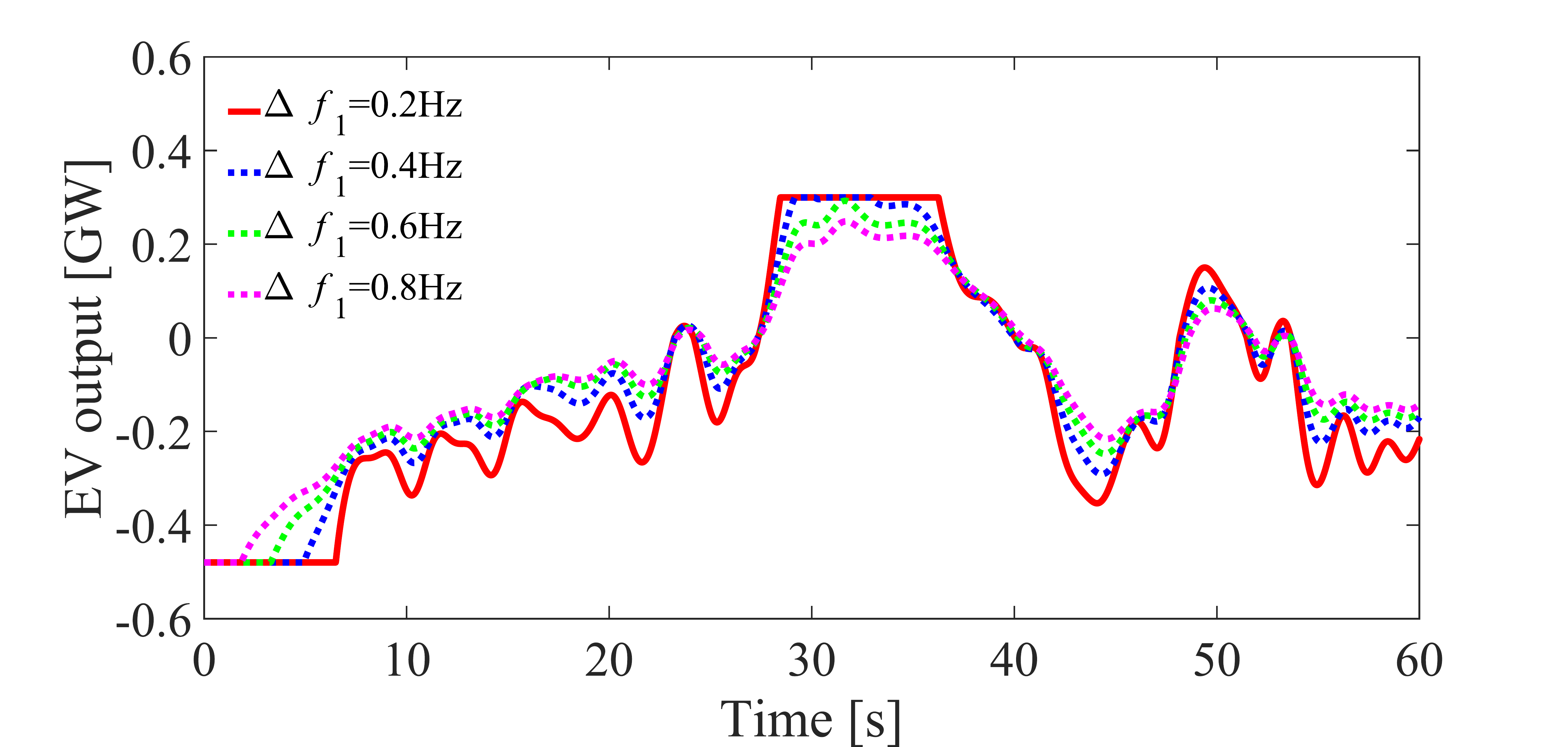} %{./figsPt1/EV2outputf1.eps}
\includegraphics[width=\hsize]{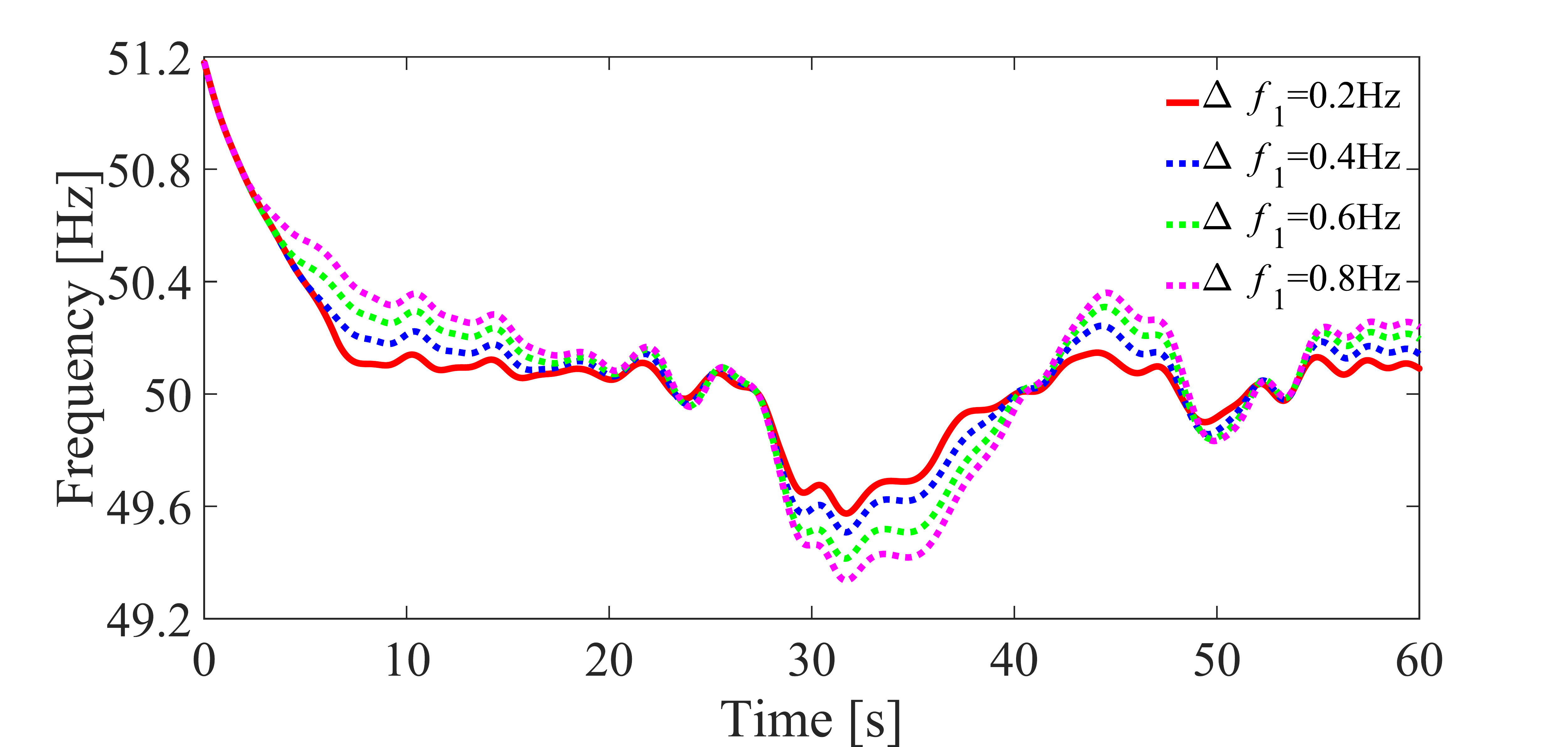} %{./figsPt1/f2f1.eps}
\subcaption{%
Whole
}%
\label{fig:Pref=0}
\end{minipage}
\begin{minipage}{.495\hsize}
\centering
\includegraphics[width=\hsize]{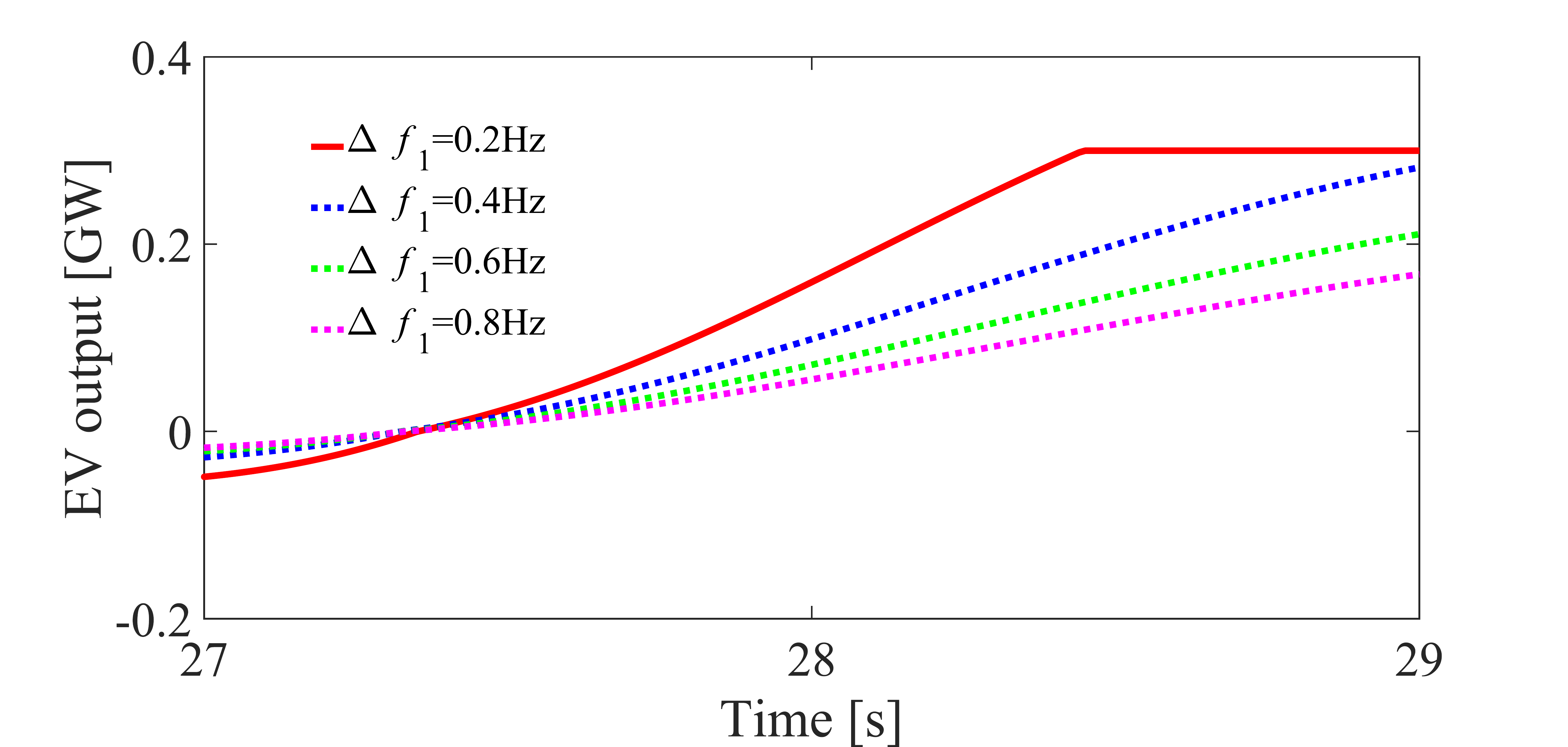} %{./figsPt1/EV_limit.eps}
\includegraphics[width=\hsize]{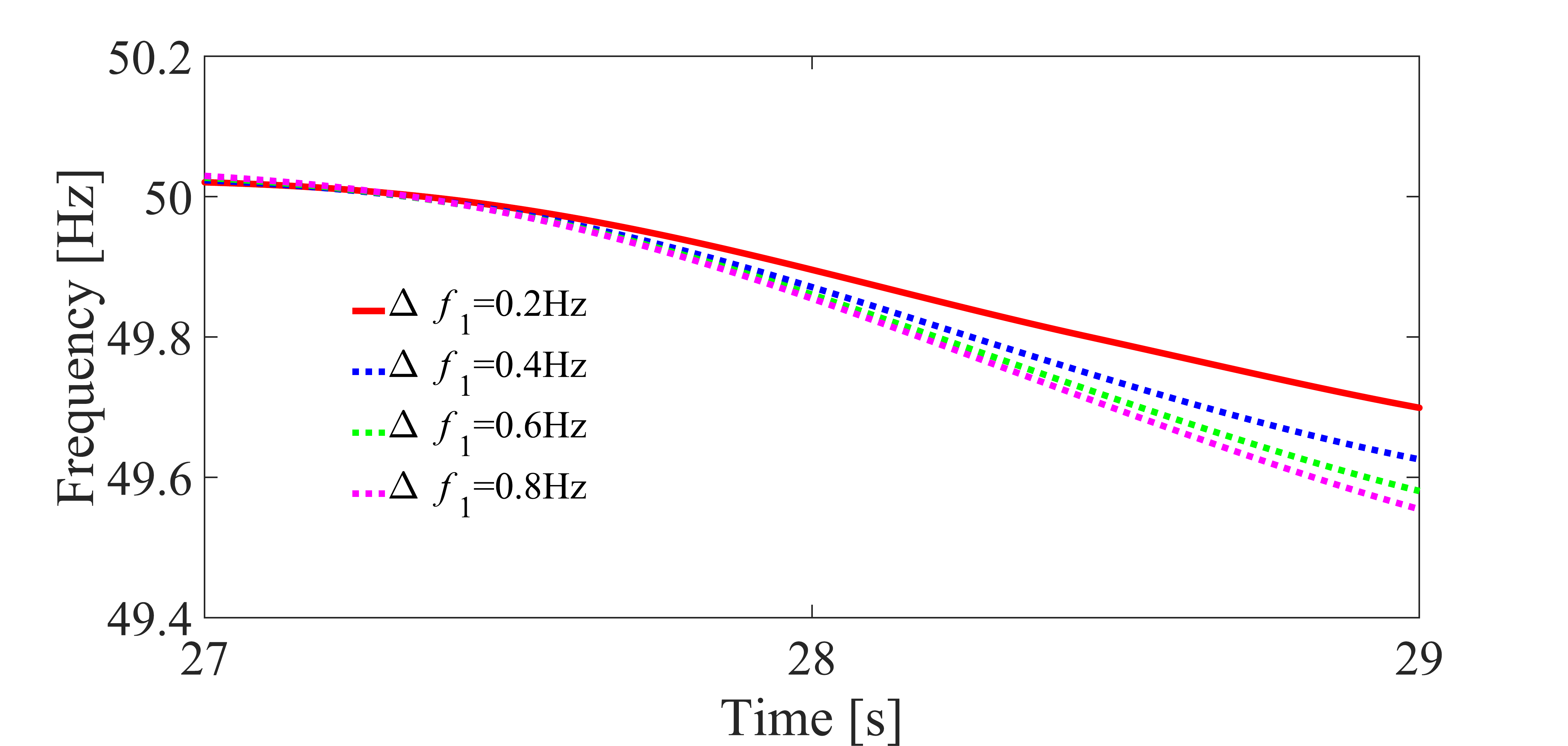} %{./figsPt1/f_limit.eps}
\subcaption{%
Zoom-up of $[27\,\U{s},29\,\U{s}]$ in (a)
}%
\label{fig:Pref=5}
\end{minipage}
\caption{%
Parameter ($\Delta f_1$) dependence of total output power from EVs and associated frequency responses under $(dV_\mathrm{cha, limit}, dV_\mathrm{discha, limit})=(80\,\U{V},\,50\,\U{V})$. 
}%
\label{fig:demo3}
\end{figure*}

For the evaluation, direct numerical simulations of the nonlinear ODE \eqref{eqn:ode} are conducted. 
Using the same manner as in \cite{Yumiki}, to evaluate $dv(L)$ for the model of multiple feeders in Figure~\ref{fig:model2}, we focus on the horizontal straight-line feeder with length $L=4.63\,\U{km}$ and measures the voltage amplitude at its end point as the rightmost part of the model. 
The estimation is done with $N_\mathrm{sta}=8$. 
In this simulation, we set $dV_\mathrm{limit}$ at the two conditions: $(dV_\mathrm{cha, limit}, dV_\mathrm{discha, limit})=(80\,\U{V},\,50\,\U{V})$ and $(80\,\U{V},\,30\,\U{V})$. 
For the frequency control, we set $\Delta f_{1}$ at $0.2\,\U{Hz}$ at first and will change its value for the performance evaluation.  
It is also assumed that there are no charging/discharging restrictions considering SoC. 
This assumption is relevant for the current time scale observed by numerical simulations. 

For comparison, we consider the two cases for the regulation of active power for the PFC reserve. 
One case is based on the proposed method; namely, distributed EVs at each station can charge (or discharge) up to the upper bound $\{-\alpha_\mathrm{cha} P_{\mathrm{EVs},i}^\mathrm{max} : i=1,\ldots,N_\mathrm{sta}\}$ (or $\{\alpha_\mathrm{discha} P_{\mathrm{EVs},i}^\mathrm{max} : i=1,\ldots,N_\mathrm{sta}\}$). 
The other case is based on the method of \cite{Ota:2012}.  
In this case, distributed EVs at each station can charge (or discharge) up to their maximum power, to which we will refer as \emph{single-objective AS}.

%%%
\subsection{Simulation Results}
\label{sec:result_dhp6}

Figure~\ref{fig:demo1} shows the numerical evaluation of the proposed design under $\Delta f_{1}=0.2\,\U{Hz}$. 
The top two figures in Figure~\ref{fig:demo1} show the upper bounds (\emph{blue} lines) at each station. 
The group of EVs at each station charges and discharges in the range (\emph{blue} line) so that the deviation of voltage amplitude at the end point of the feeder can be within the upper acceptable limit. 
In the figures (a) and (b), we set $dV_\mathrm{limit}$ at the two conditions: $(dV_\mathrm{cha, limit}, dV_\mathrm{discha, limit})=(80\,\U{V},\,50\,\U{V})$ and $(80\,\U{V},\,30\,\U{V})$. 
In the figure (a), we use the computed values $(\alpha_\mathrm{cha, limit}, \alpha_\mathrm{discha, limit})=(0.7335, 0.4585)$, in the figure (b) we use $(\alpha_\mathrm{cha, limit},$ $ \alpha_\mathrm{discha, limit})=(0.7335, 0.2751)$.  
The middle two figures show the results on total active power output by EVs. 
The \emph{red, solid} line shows the active power output for single-objective AS, while the \emph{blue, dashed} line does the active power output for multi-objective AS. 
The bottom two figures show the time responses of frequency of the transmission grid. 
The responses are originally caused by the time-varying PV generation and load consumption.\footnote{The deviation is relatively large by comparison with standard bounds of the grid's frequency: $[50\U{Hz}-0.2\U{Hz},50\U{Hz}+0.2\U{Hz}]$ for eastern Japan. 
This is mainly because as stated in Section~\ref{sec:Simulation Setting}, the introduction rate of PV is large in the current setting, and its smoothing effect is not considered.} 
The \emph{black, solid} line shows the time response of frequency without EVs, while the \emph{red, solid} (or \emph{blue, dashed}) line does the time responses for single-objective (or multi-objective) AS. 
By comparison of the frequency responses with/without EVs, we see that the frequency approaches the nominal value, $50\,\U{Hz}$, by charging/discharging from EVs. 
Here, we compare the frequency responses with EVs for single- and multi-objective AS. 
The \emph{red, solid} line (single-objective AS) is closer to the nominal value than the \emph{blue, dashed} line (multi-objective AS). 
For this, the charging/discharging from EVs at $[0\,\U{s}, 7\,\U{s}]$ and $[28\,\U{s}, 36\,\U{s}]$ in the middle of Figure~\ref{fig:demo1}(a) (or $[39\,\U{s}, 53\,\U{s}]$ in the middle of Figure~\ref{fig:demo1}(b)) is bounded with the proposed design. 
This implies that the total amount of supply power by EVs in multi-objective AS is smaller than that in single-objective AS, and that the single-objective AS shows a better performance for the frequency control as expected.  

The associated data on distribution voltage for the numerical evaluation are shown in Figure~\ref{fig:demo2}. 
The voltage amplitude was computed with the nonlinear ODE \eqref{eqn:ode}. 
The top two figures show the time series of distribution voltage at $x=L$ in Figure~\ref{fig:model2}. 
The two \emph {black, dashed} lines show the upper/lower limits of voltage determined by $dV_\mathrm{cha, limit}$ or  $dV_\mathrm{discha, limit}$. 
The \emph{red, solid} (or \emph {blue, dashed}) lines shows the time series of distribution voltage for single-objective (or multi-objective) AS. 
The figure (a) (or (b)) indicates that the voltage deviation at the end point of feeder is mitigated by considering the upper bound with $\alpha_\mathrm{cha}$ (or $\alpha_\mathrm{discha}$). 
The bottom two figures show the spatial profiles of distribution voltage along the horizontal straight-line feeder in  Figure~\ref{fig:model2}. 
The figure (a) shows the voltage profile at time $2\,\U{s}$, and the figure (b) shows the voltage profile at $40\,\U{s}$. 
In the figure (a), the \emph{solid black} line shows the voltage for the loads, the \emph{red, solid} (or \emph{blue, dashed}) line shows that for both the loads and EVs at the charging mode with unregulated $-P_{\mathrm{EVs}, i}^\mathrm{max}$ (or regulated $-\alpha_\mathrm{cha}P_{\mathrm{EVs}, i}^\mathrm{max}$). 
In the figure (b), the \emph{solid black} line shows the voltage drop by the loads, the \emph{red, solid} (or \emph{blue, dashed}) line shows that for both the loads and EVs at discharging mode with unregulated $P_{\mathrm{EVs}, i}^\mathrm{max}$ (or regulated $\alpha_\mathrm{discha} P_{\mathrm{EVs}, i}^\mathrm{max}$). 
From the figures (a) and (b), we see that the distribution voltage profile is regulated to be within the determined range by the proposed design. 
Here, the deviation estimated with the nonlinear ODE \eqref{eqn:ode} were $83.5\,\U{V}$ in the figure (a) and $29.4\,\U{V}$ in the figure (b), which were slightly different from the fixed $dV_\mathrm{cha, limit}$ and $dV_\mathrm{discha, limit}$. 
This is mainly because the proposed design is based on the approximate solution of the nonlinear ODE \eqref{eqn:ode}, which will be discussed in the next subsection. 

%%%
\subsection{Discussion}

In the above subsection we showed the performance of the proposed autonomous V2G design numerically. 
The capability for providing the PFC reserve to TSO is evaluated with the measures in \eqref{eqn:reserve}. 
As mentioned above, the capability depends on the choice of the parameter $\Delta f_1$. 
Thus, the parameter dependence of the total output power from EVs and associated frequency responses are shown in Figure~\ref{fig:demo3}. 
The value of $\Delta f_{1}$ varies from $0.2\,\U{Hz}$ to $0.8\,\U{Hz}$. 
This figure shows that the PFC reserve is consistently provided for the different values of the parameter $\Delta f_1$.    
In Figure~\ref{fig:demo3}(b), we see that the frequency approaches the nominal value rapidly as the value of $\Delta f_{1}$ becomes smaller. 
This is clearly characterized by the capability measure $\Delta P_\mathrm{discha/Hz }$ in \eqref{eqn:reserve}, which is inversely proportional to $\Delta f_1$. 
Hence, the frequency characteristic requirement for the PFC reserve (if posed from TSO) can be fulfilled sufficiently for a small choice of $\Delta f_{1}$. 

\begin{figure}[t]
\centering
\includegraphics[width=.675\textwidth]{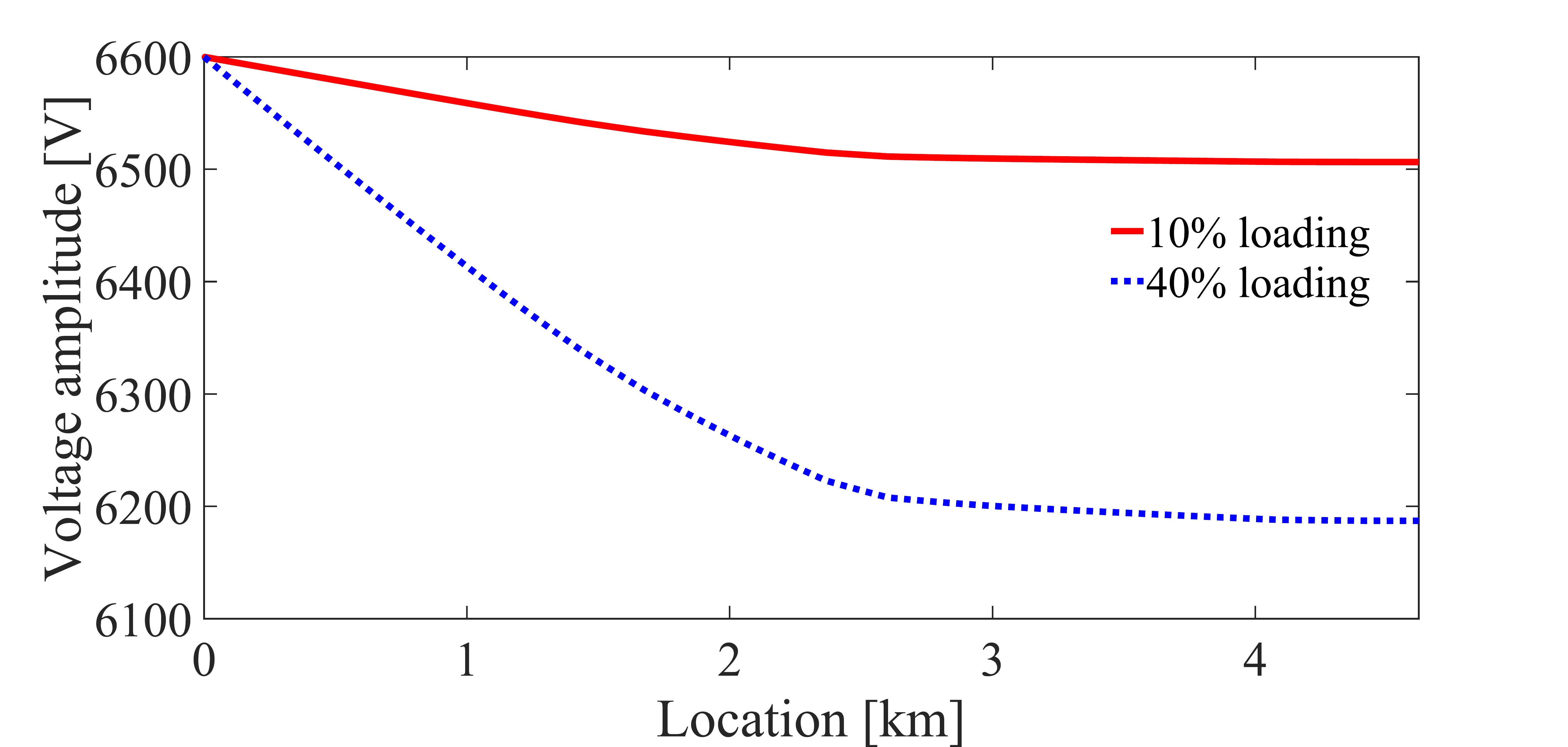} %{./figs/verror.eps}
\caption{%
Numerical results on distribution voltage profiles for 10\% and 40\% of the loading in Figure~\ref{fig:model2}.
}%
\label{fig:verror}
\centering
\vspace*{2mm}
\includegraphics[width=.6\textwidth]{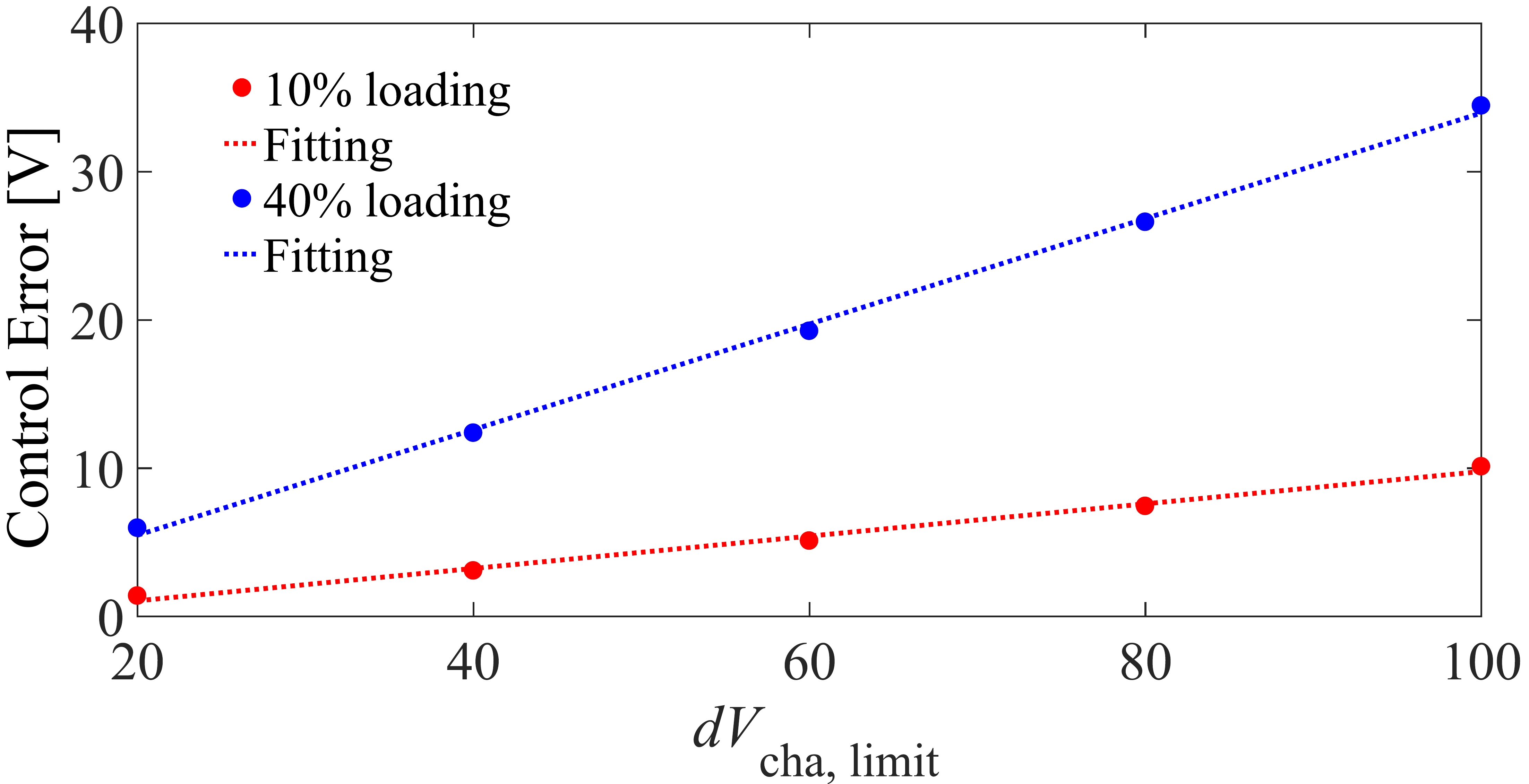} %{./figs/Fitting_rev.eps}
\caption{%
Numerical results on control error for the direct numerical simulations in Figure~\ref{fig:model2}.
}%
\label{fig:error}
\end{figure}

In the end of the last subsection, we mentioned a small error for the voltage regulation in the proposed design. 
The control is parameterized with the parameter $dV\sub{limit}$ that is assigned by DSO. 
Here, we discuss about how the control error depends on the choice of $dV_\mathrm{limit}$ and loading condition of the distribution feeder.    
The total amount of the loads is set as the two conditions: 10\% and 40\% of the bank's rated capacity in Figure~\ref{fig:model2}. 
The associated direct numerical results on distribution voltage profiles are presented in Figure~\ref{fig:verror}, where the \emph{red, solid} (or \emph{blue, dashed}) line shows the voltage for 10\% loading (or 40\% loading). 
Note that the voltage drop stops in the part of the feeder after 2\,km because the loading center is located in the feeder before 2\,km. 
Here we define the control error as the absolute value of the difference between the direct simulation at the end of feeder and  $dV_\mathrm{cha, limit}$. 
The $dV_\mathrm{cha, limit}$-dependence of the control error is shown in Figure~\ref{fig:error}. 
In Figure~\ref{fig:error}, for each of the loading conditions, the control error increases with $dV_\mathrm{cha, limit}$. 
This basically comes from the approximation contained in the ODE-based modeling---the fact that the approximate solution of the nonlinear ODE \eqref{eqn:ode} is derived under the condition that all the voltage amplitudes $v$ on the right-hand side of (\ref{eqn:ode}) are close to unity. 
Here, we can see in Figure~\ref{fig:error} that the control error behaves in a linear manner with $dV_\mathrm{cha, limit}$ and the loading condition.  
This is helpful because the control error is predictable with the loading condition and can be hence compensated in the application.

%%%%
%%%%
\section{Power-HIL (Hardware-In-the-Loop) Testbed}
\label{sec:HIL}

This section is a brief review of the development of Power-HIL testbed for experimental validation of the autonomous V2G design for multi-objective AS. 
This is built as a laboratory facility of a real-time digital simulator and physical devices including an EV battery. 

%%%
\subsection{Overview}

\begin{figure}[t]
\centering
\includegraphics[width=.7\textwidth]{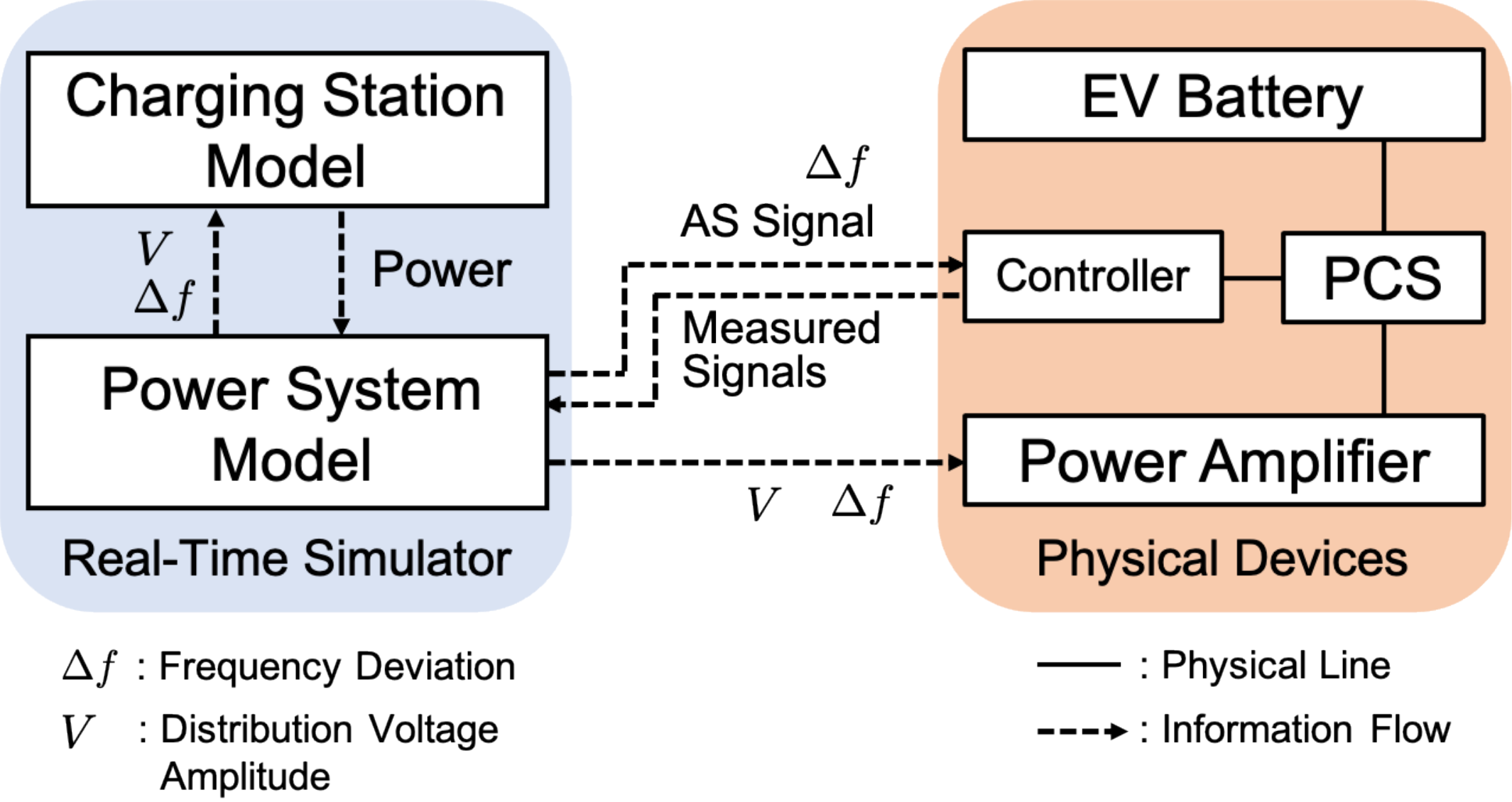} %eps}
\caption{%
Overview of Power-HIL (Hardware-In-the-Loop) testbed built in this paper.  
Digital and analog components in this testbed are connected via power and communication lines.  
}%
\label{fig:Overview}
\end{figure}

First, we describe the Power-HIL testing for this study. 
Figure~\ref{fig:Overview} shows the overview of Power-HIL testbed for validating the multi-objective AS. 
The testbed includes a real-time digital simulator and physical devices of PCS, controller, power amplifier, and EV battery. 
In the real-time digital simulator, the frequency dynamics of a transmission grid and the voltage dynamics of a distribution grid are simulated with mathematical models, denoted by ``Power System Model" in Figure~\ref{fig:Overview}. 
That is, the frequency deviation $\Delta f$ and the distribution voltage at the stations are calculated with the models introduced in Section \ref{subsec:PSM} and sent to ``Charging Station Model" in Figure~\ref{fig:Overview}. 
The block ``Charging Station Model" computes the AS signals for the stations according to the control logic \eqref{eqn:droop2} in Section~\ref{sec:autonomou}, which provides the regulation law of active power output for the provision of multi-objective AS. 
In addition to the digital simulator, the AS signal is sent from ``Charging Station Model " through a communication line to ``Controller" in Figure~\ref{fig:Overview}. 
The block ``Controller" is used for commanding the AS signal (including the information on $\Delta f$) to ``PCS," namely ``EV Battery." 
The Power-HIL testbed is therefore closed in loop by sending the signal to ``PCS,"  measuring its actual value of charging/discharging power, and receiving the measured signal at ``Power System Model."  
In addition, the frequency deviation $\Delta f$ and the distribution voltage in ``Power System Model" are sent to ``Power Amplifier" that simulates in the physical domain the temporal change of distribution voltage at one station where the real ``EV Battery" is connected.  

%%%
\subsection{Power System Model}
\label{subsec:PSM}

\begin{figure}[t]
\centering
\includegraphics[width=.6\textwidth]{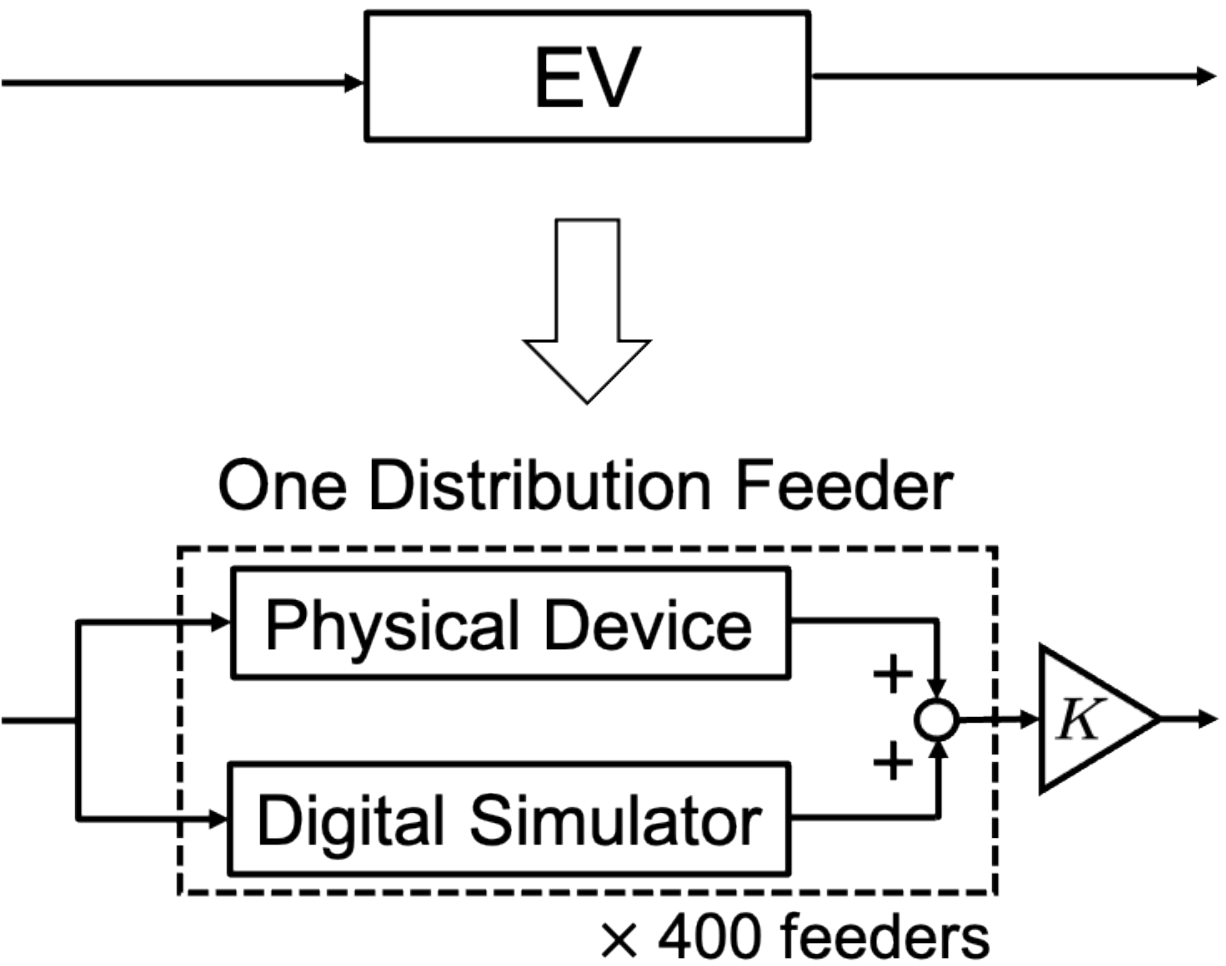} %eps}
\caption{%
Change of the block diagram in Figure~\ref{fig:f} for Power-HIL testing. 
It contains multiple distribution grids connected to electric vehicles that provide the multi-objective ancillary service.
}%
\label{fig:Trans_HIL}
\centering
\includegraphics[width=.6\textwidth]{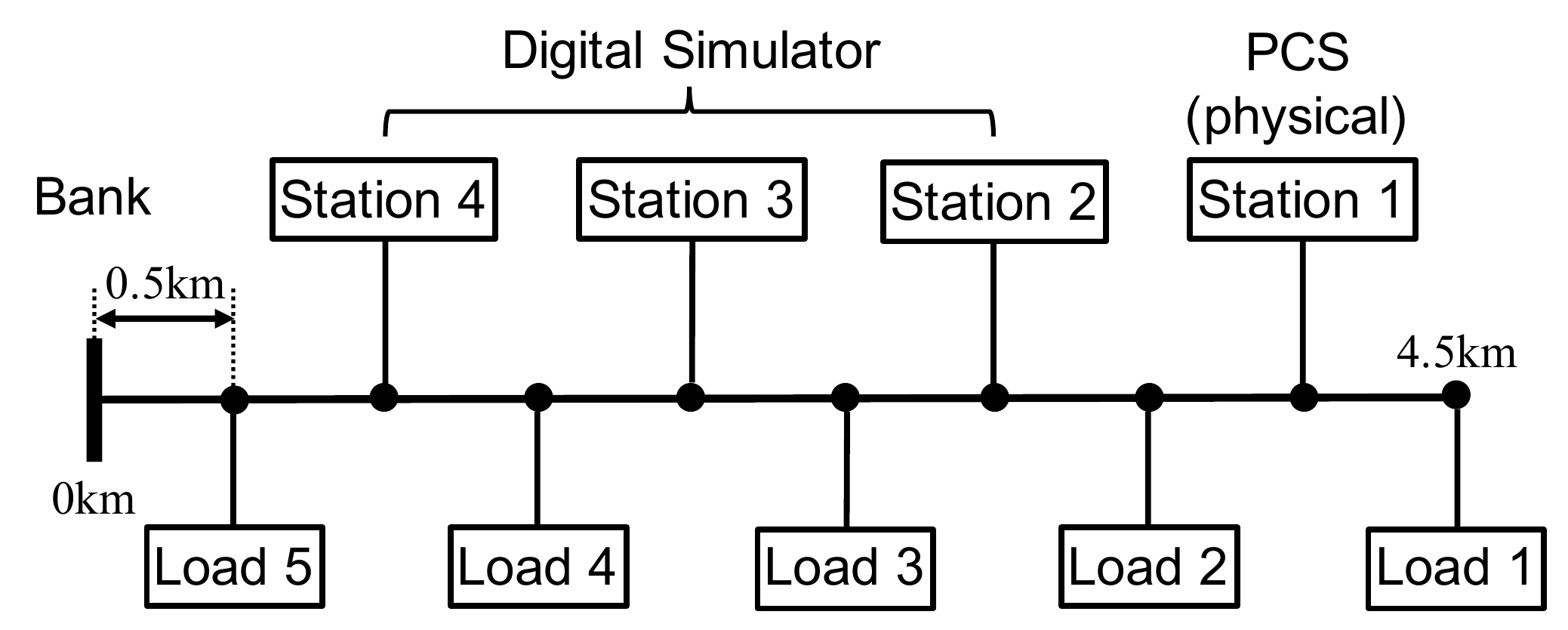} 
\caption{%
Model of one distribution feeder used for experimental validation. 
The first charging station ``Station~1'' is simulated with Power Conditioning System (PCS), and the other stations with the real-time digital simulator. 
}%
\label{fig:feeder_HILS}
\end{figure}

\begin{figure}
\centering
\includegraphics[width=.6\textwidth]{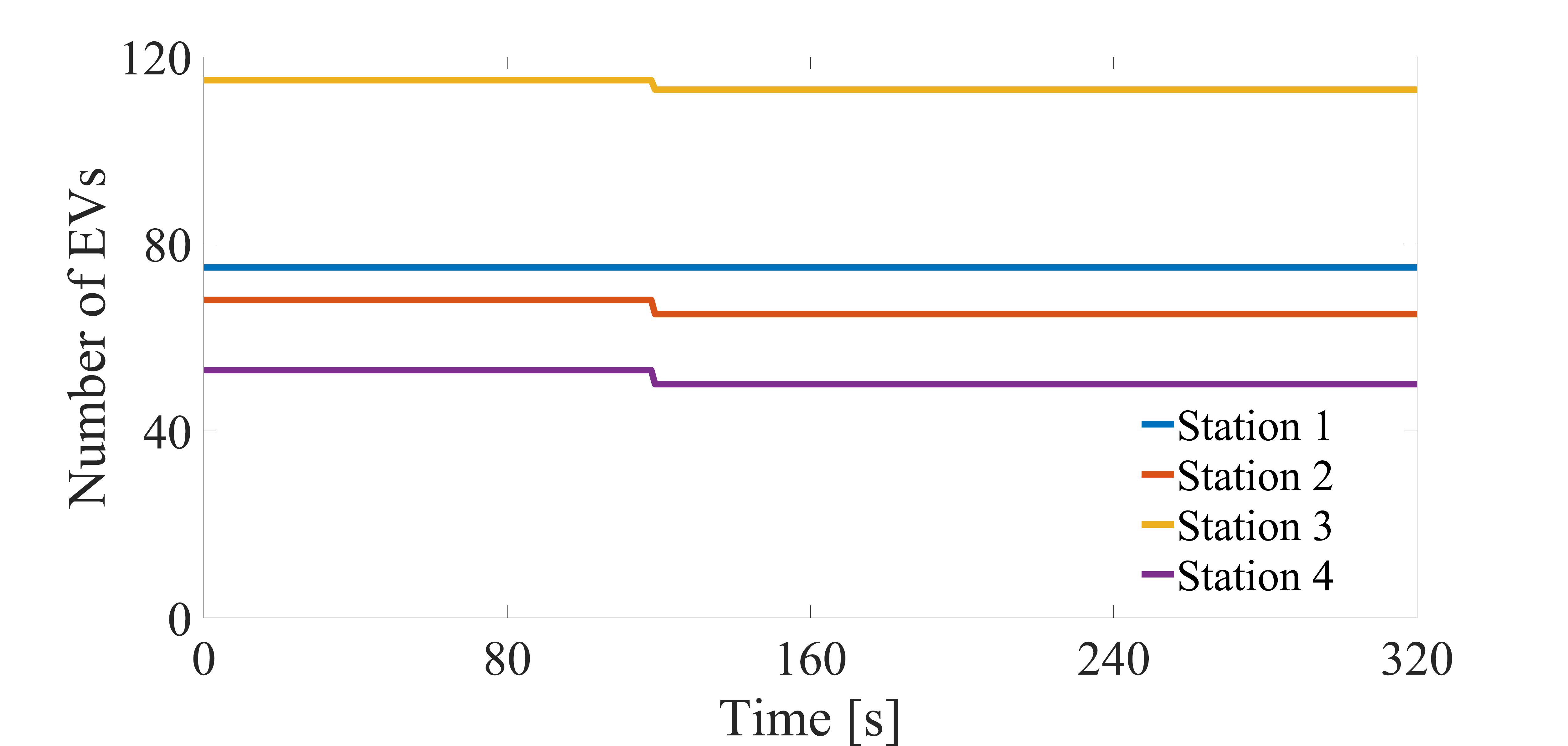} %{./figsPt2/Number_HILS2.eps}
\caption{%
Temporal change of the number of electric vehicles (EVs) at the 4 charging stations in the distribution feeder of Figure~\ref{fig:feeder_HILS}. 
}%
\label{fig:Number}
\end{figure}

In this section, we describe the mathematical models of transmission and distribution grids denoted by ``Power System Model" in Figure~\ref{fig:Overview}. 
The setting of the models and parameters is basically from \cite{Mizuta:2019}. 
Figure~\ref{fig:f} is used for the model of transmission grid, where the block ``EV" is replaced with the block model of multiple distribution grids as shown in Figure~\ref{fig:Trans_HIL}. 
It is here assumed that 400 same feeders of the distribution grid are connected to the transmission grid. 
Figure~\ref{fig:feeder_HILS} shows the model of one distribution feeder that is a single, straight-line feeder with length $4.5\,\U{km}$.  
The setting is used in \cite{Mizuta:2019}. 
The feeder has the five loads and four charging stations located at a common interval $0.5\,\U{km}$. 
All electrical components in the distribution grid are modeled with Sim Power Systems in MATLAB/Simulink including voltage sources in the distribution substation, distribution lines, pole transformers, customer loads, and charging stations. 
The consumer loads and the charging stations are modeled as constant power sources.  
Electrical transients for inductors of lines and transformers are also considered. 
This is why dynamical characteristics of the distribution grid can be incorporated, unlike the nonlinear ODE model in Section~\ref{sec:exp} that represents a steady profile of distribution voltage.  
As the contribution from transportation, we use synthetic operation on the number of EVs at each charging station shown in Figure~\ref{fig:Number}, where each EV has rated power output of $4\,\U{kW}$. 
This is because the maximum active power output of the real EV battery we use in the Power-HIL simulation is $6\,\U{kW}$ described as in next section. 
Figure~\ref{fig:Number} is based on the practical data in the EV-sharing demonstration project and used in Section~\ref{sec:exp}.  
The loads in the distribution feeder are represented as the constant power model: $350\,\U{kW}$ for Load1 and Load3; $300\,\U{kW}$ for Load2, Load4, and Load5. 
In Figure~\ref{fig:feeder_HILS}, ``Station~1" is built as PCS, and the other stations are built on the digital simulator. 

\begin{figure}[t]
\centering
\includegraphics[width=.6\textwidth]{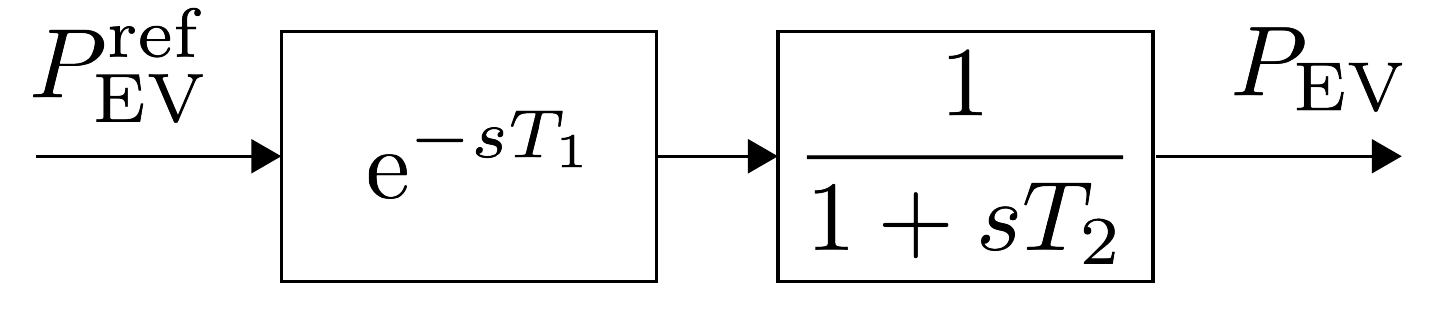}
\caption{%
Dynamic model of input/output response for power conversion. 
The input is the ancillary-service (AS) signal $P_\mathrm{EV}^\mathrm{ref}$, and the output is the active-power output $P_\mathrm{EV}$ of PCS. 
The model is adopted in Figure~\ref{fig:feeder_HILS} to the three stations built in the digital simulator. 
}%
\label{fig:delay}
\end{figure}

\begin{figure*}[t]
\centering
\includegraphics[width=.95\textwidth]{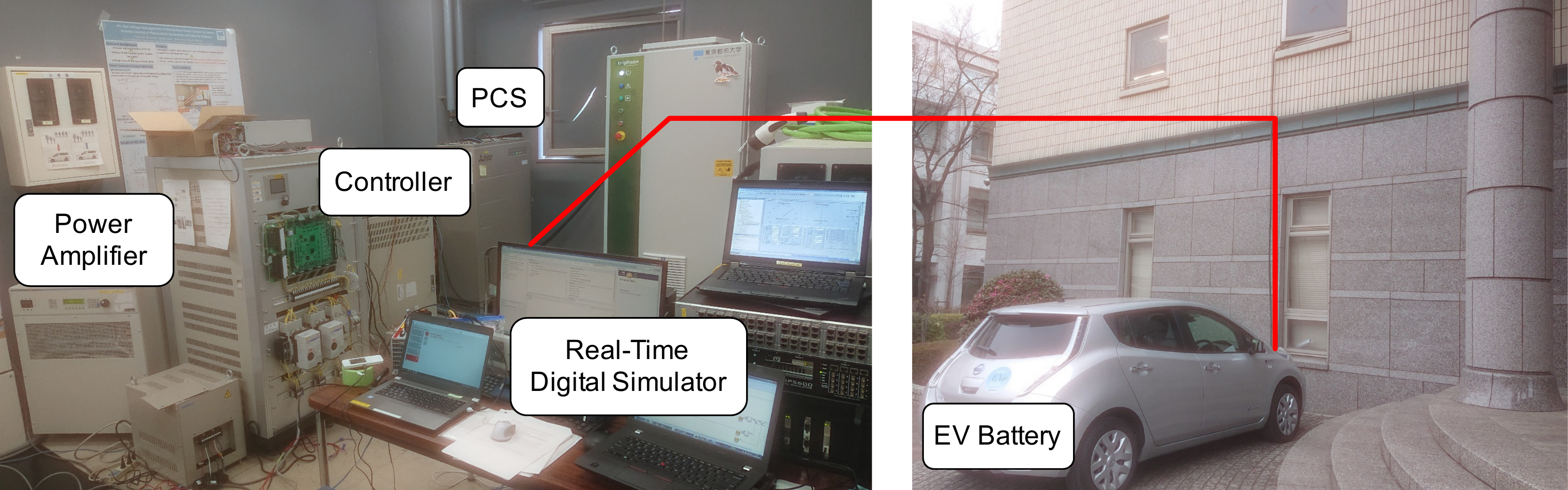}
\caption{%
Configuration of Power-HIL (Hardware-In-the-Loop) testbed. 
It contains the real-time digital simulator OPAR-RT Technologies (Model \#OP5600), the power amplifier AMETEK (Model \#MX-15-1pi), the controller (dSPACE, Micro Auto Box II), and the Power Conditioning System (PCS) (Mitsubishi Electric Corporation, Model \#EVP-SS60B3-M7-R), EV battery (Nissan Motor Co., Ltd, Nissan Leaf). 
The left picture shows the experimental setup of Power-HIL, and the right picture shows the EV battery. 
The EV battery in the right picture is connected to the PCS in the left picture as shown by the red line between the two pictures. 
}
\label{fig:lab}
\end{figure*}

Finally, we consider latency in the real system, precisely time-delays due to the local measurement and communication in ``Controller" of Figure~\ref{fig:Overview} and a time-lag inherent in the physical PCS. 
Their consideration does not appear in the preceding paper \cite{Mizuta:2019}. 
A time-delay can happen when ``Controller" in Figure~\ref{fig:Overview} receives the AS signal from ``Power System Model" and commands it to ``PCS" and ``EV battery." 
In addition to the measurement, because the multi-objective AS is intended to a practical large-scale grid, another time-lag due to long-distance communication is inevitable and should be taken into account for the feasibility testing. 
Furthermore, the dynamics of the grid's frequency develop in time scale of several seconds, and the PFC reserve by EVs is thus intended to work in the same time scale. 
The time scale is also involved in dynamic characteristics of the PCS, precisely speaking, time-lag from input to output in the PCS. 
The digital simulator without consideration of time-delay and time-lag is described in \cite{Kamo:2018}. 
To incorporate these effects in the digital simulator, we use a simple-to-implement model in Figure~\ref{fig:delay} for the input/output response of power conversion. 
The input is the AS signal $P_\mathrm{EV}^\mathrm{ref}$, and the output is the active-power output $P_\mathrm{EV}$ of PCS. 
The model is adopted in Figure~\ref{fig:feeder_HILS} to the three stations built in the digital simulator.  
In this study, we set the parameter $T_{1}$ of time-delay as $0.30\,\U{s}$ and $T_{2}$ of time-lag as $0.43\,\U{s}$.  
These values are based on the measurement of physical devices, which will be discussed in Figure~\ref{fig:compare_inout} through comparison. 
It should be remarked that the simple-to-implement model is a main challenge to the real-time platform. 
For the platform, conventional, full models of power-electronics-interfaced equipments, which are multi-scale dynamical systems with many nonlinearities, are hopeless. 

Here, it is noted that no voltage regulation device is considered in the following analysis because the effectiveness of the proposed autonomous V2G design is evaluated in a relatively simple setting. 
Related to this, the time-varying PV generation is considered in the model of transmission grid not the distribution one. 

%%%
\subsection{Experimental Setup}

Figure~\ref{fig:lab} shows the two pictures of the Power-HIL testbed. 
The testbed contains the real-time digital simulator manufactured by OPAL-RT Technologies (Model \#OP5600) and located in the lower part of the left picture in Figure~\ref{fig:lab}. 
The time step for the digital simulator is $100\,\U{\mu s}$ for calculation of the frequency deviation, the distribution voltage, and the charging/discharging power for the charging station. 
The setting of the time step is enough to the tractable simulation of the distribution grid that exhibits the fastest transient phenomenon in this testing. 
The frequency deviation $\Delta f$ is calculated every time step, namely $100\,\U{\mu s}$. 
Figure~\ref{fig:lab} also shows the configuration of Power-HIL. 
The power amplifier is manufactured by AMETEK (Model \#MX15-1pi; Rated AC output power is $15\,\U{kVA}$), which is located in the left part of the left picture in Figure~\ref{fig:lab}, and is used for the physical simulation of the frequency dynamics and the distribution voltage.  
The power amplifier outputs the distribution voltage to PCS that is manufactured by Mitsubishi Electric Corporation (Model \#EVP-SS60B3-M7-R; 6kVA) in the upper part of the left picture in Figure~\ref{fig:lab} and is used for the physical simulation of charging/discharging of ``Station~1" in Figure~\ref{fig:feeder_HILS}. 
The equipment of ``Controller" is manufactured by dSPACE (Micro Auto Box II), and the EV battery is manufactured by Nissan Motor Co., (Model Nissan Leaf; $30\,\U{kWh}$). 

We here set the values of $dV_\mathrm{cha, limit}$ and $dV_\mathrm{discha, limit}$ at $80\,\U{V}$.   
By using the synthetic operation data on EVs in Figure~\ref{fig:Number} and the method in Section~\ref{sec:main}, the upper bound for charging/discharging patterns is pre-computed as  $\{-\alpha_\mathrm{cha} P_{\mathrm{EVs},i}^\mathrm{max} : i=1,\ldots,N_\mathrm{sta}\}$ (or $\{\alpha_\mathrm{discha} P_{\mathrm{EVs},i}^\mathrm{max}: i=1,\ldots,N_\mathrm{sta}\}$) with $N_\mathrm{sta}=4$ in Figure~\ref{fig:feeder_HILS}. 
The parameter $\Delta f_{1}$ is set at $0.2\,\U{Hz}$ from the simulation result in the last part of Section~\ref{sec:exp}.

%%%%
%%%%
\section{Feasibility Testing}
\label{sec:results}

The aim of this section is to establish the practical feasibility of the proposed autonomous V2G design. 
For this, we show a series of experimental results on the Power-HIL testing and their implications.  

%%%
\subsection{Primary Frequency Control}
\label{sec:PFCexp}

\begin{figure}[t]
\centering
\includegraphics[width=.7\textwidth]{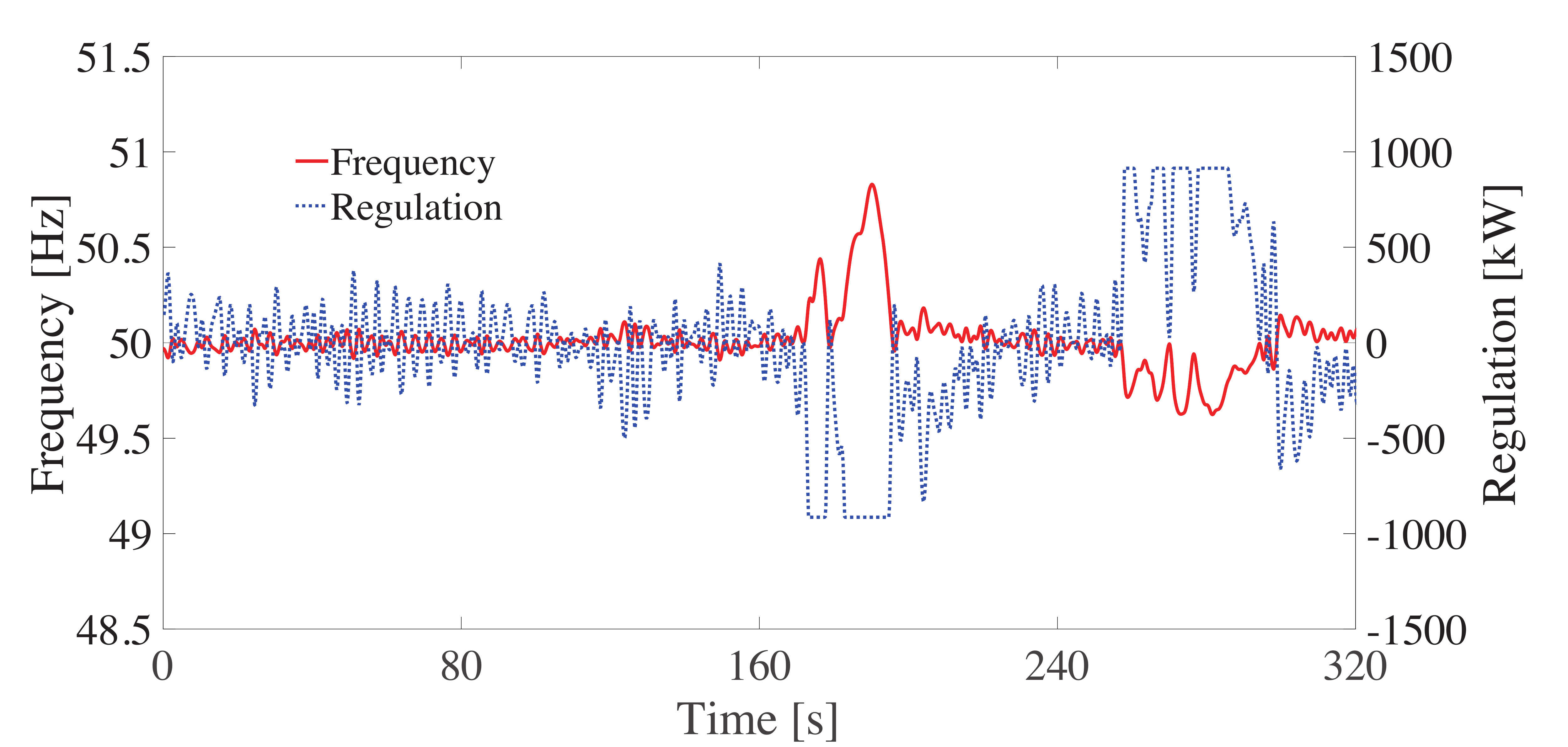} %{./figs/f_HIL.eps}
\caption{%
Generated ancillary-service (AS) signals to one distribution feeder as the primary frequency control (PFC) reserve and associated frequency responses of the transmission grid.  
The \emph{blue, dashed} line shows the AS signals, and the \emph{red, solid} line does the grid's frequency.
}%
\label{fig:compare_hils}
\end{figure}

\begin{figure*}[t]
\centering
\begin{minipage}{\textwidth}
\centering
\includegraphics[width=.49\hsize]{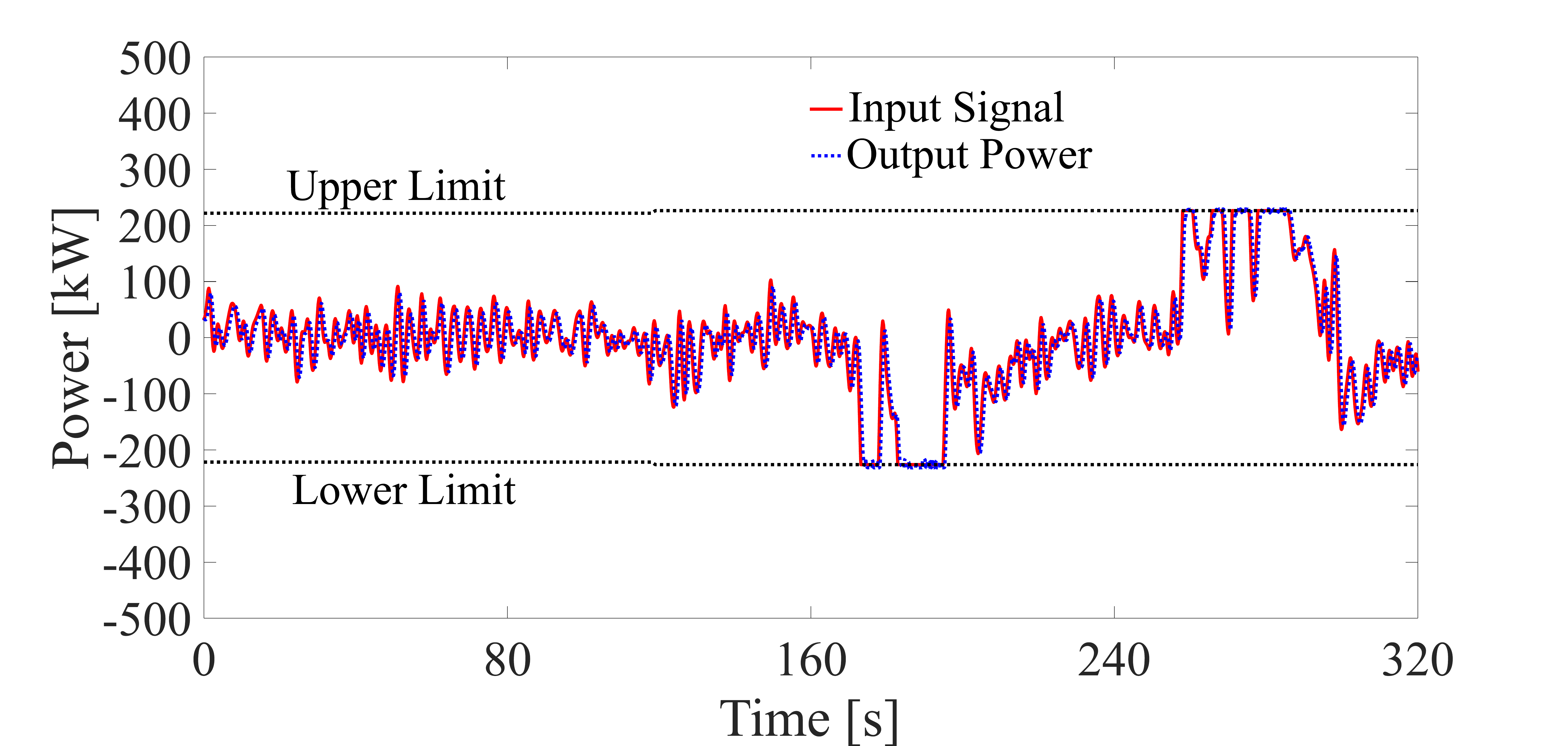} %{./figs/Sta1_limit.eps}
\includegraphics[width=.49\hsize]{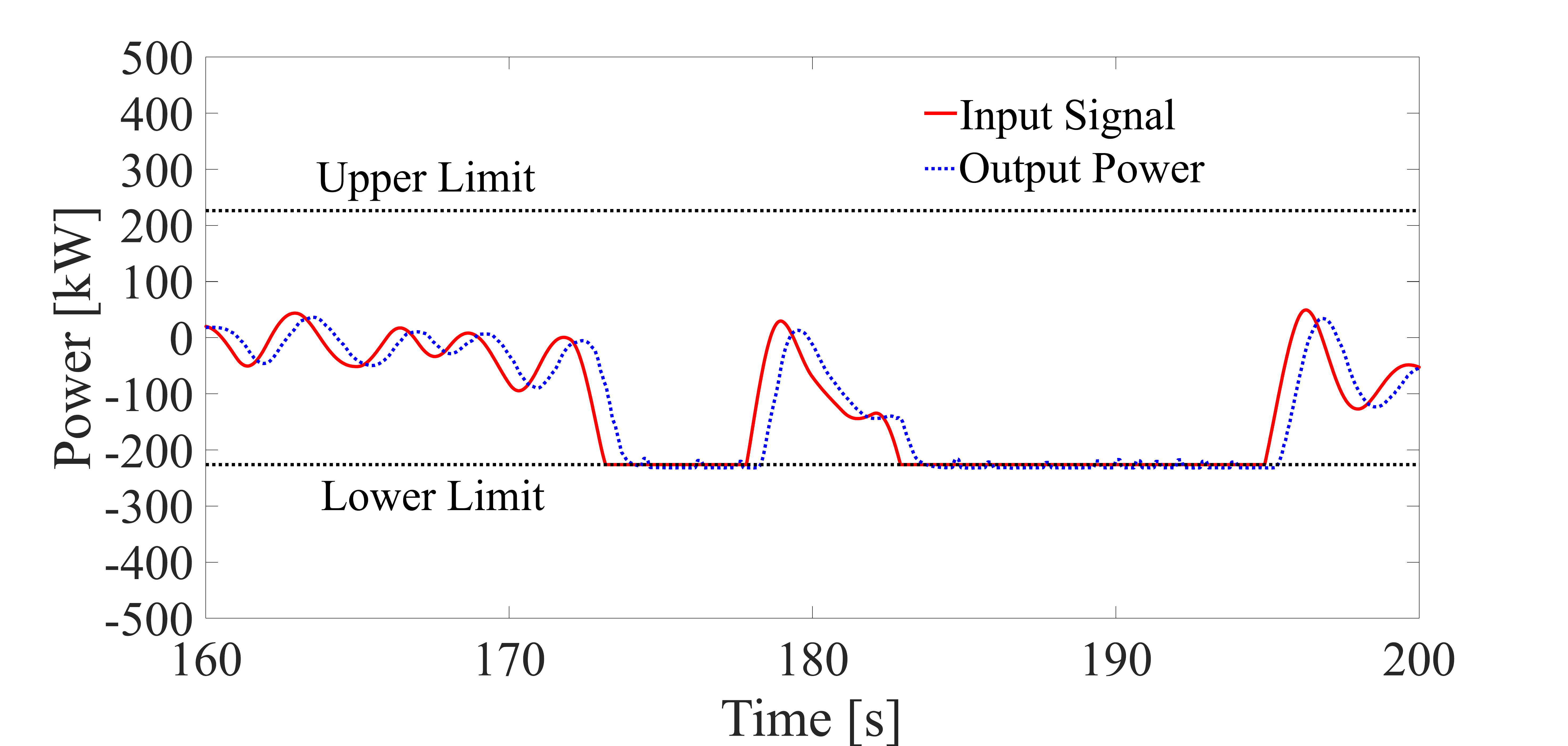} %{./figs/Sta1_limit2.eps}
\subcaption{%
\footnotesize
Station~1 (built on Power Conditioning System (PCS))
}%
\end{minipage}
\begin{minipage}{\textwidth}
\centering
\includegraphics[width=.49\hsize]{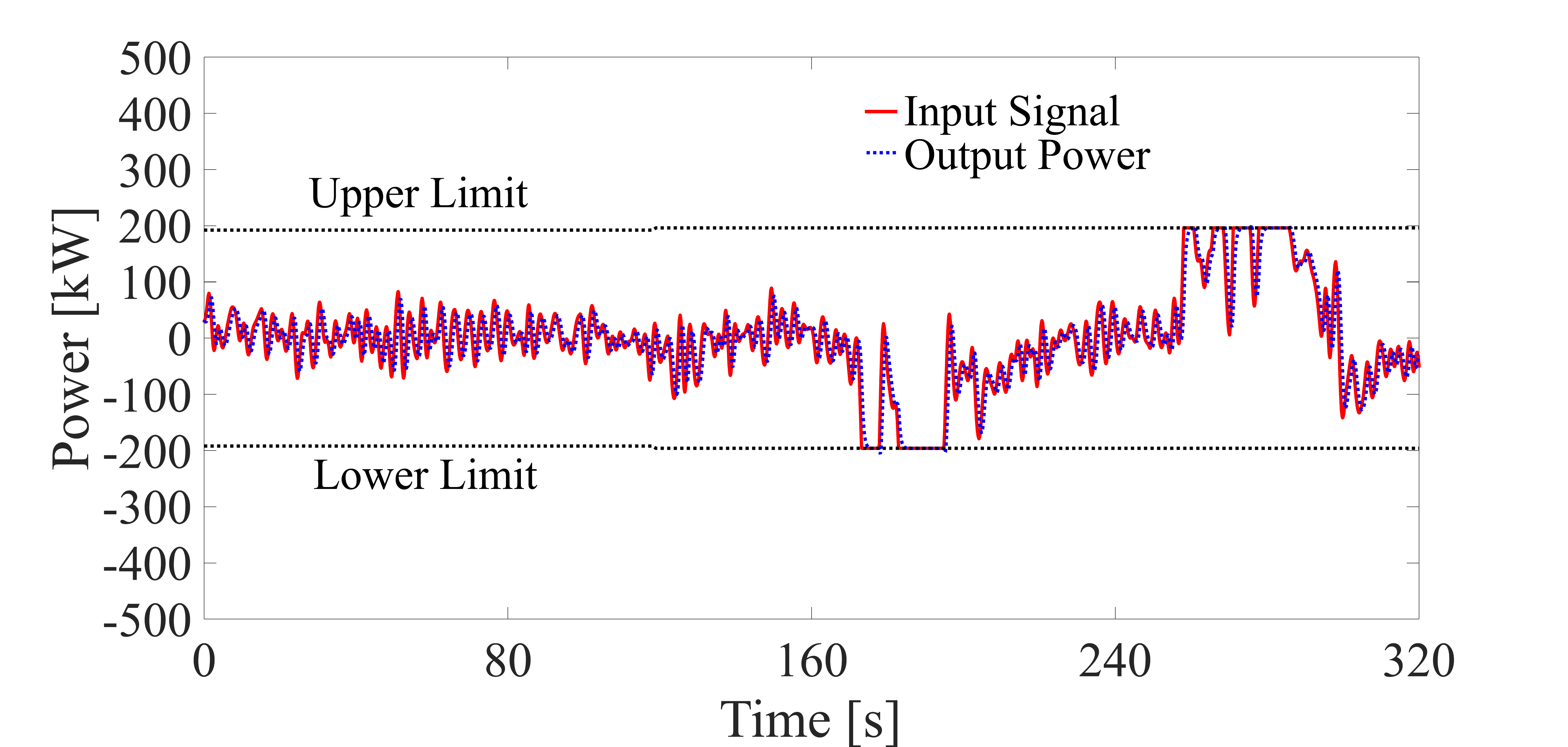} %{./figs/Sta2_limit.eps}
\includegraphics[width=.49\hsize]{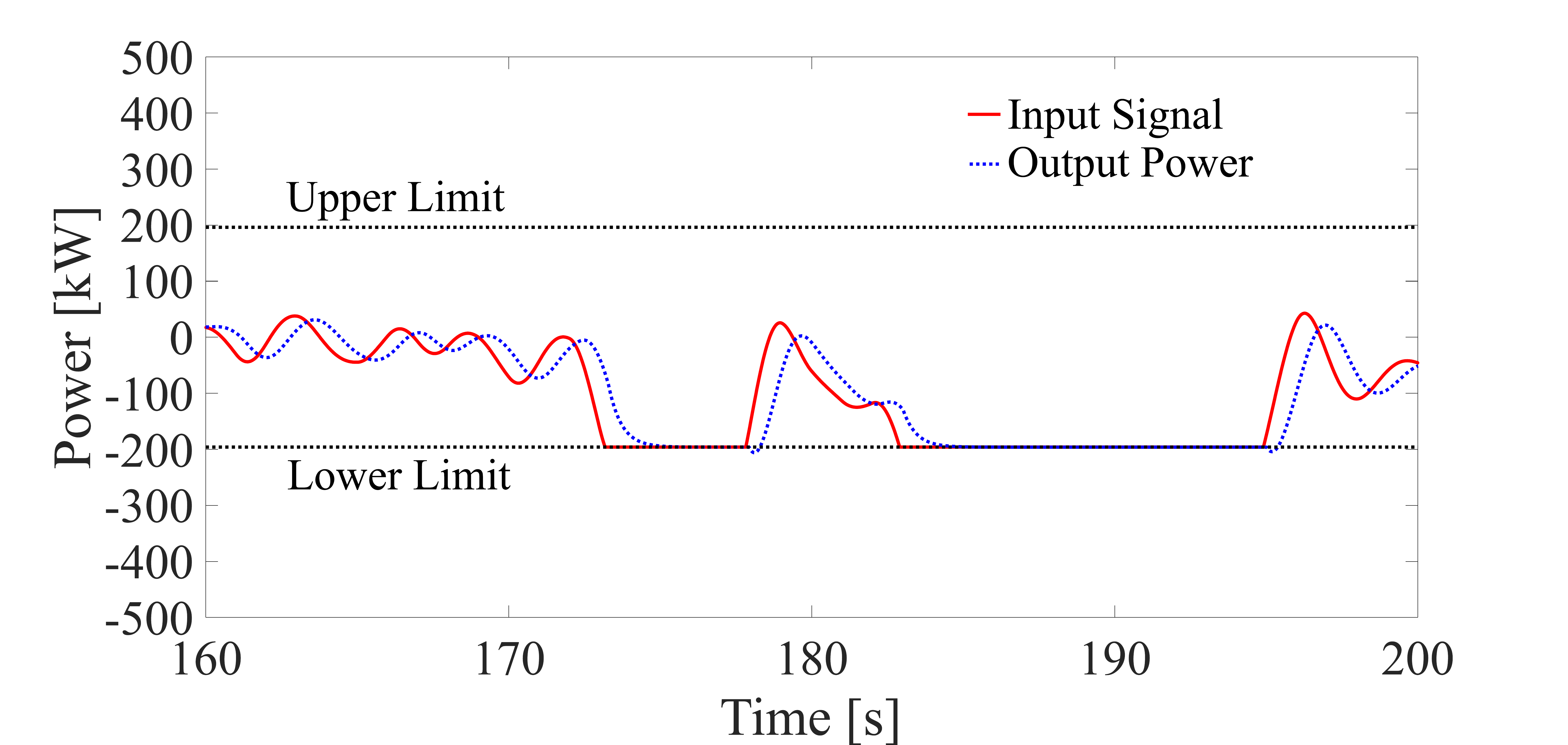} %{./figs/Sta2_limit2.eps}
\subcaption{%
\footnotesize
Station~2 (built on the digital simulator)
}%
\end{minipage}
\caption{%
Input/output responses of the two charging stations associated with Figure~\ref{fig:compare_hils}.  
The \emph{red, solid} lines show the input ancillary-service (AS) signals, and the \emph{blue, dashed} lines show the output power, and the two \emph{black, dashed} lines show determined the upper/lower limits. 
The left and right figures show the whole time-series during $[0\,\U{s}, 320\,\U{s}]$ and their zoom-up during $[160\,\U{s}, 200\,\U{s}]$, respectively. 
}%
\label{fig:compare_inout}
\end{figure*}

\begin{figure}[t]
\centering
\includegraphics[width=.7\textwidth]{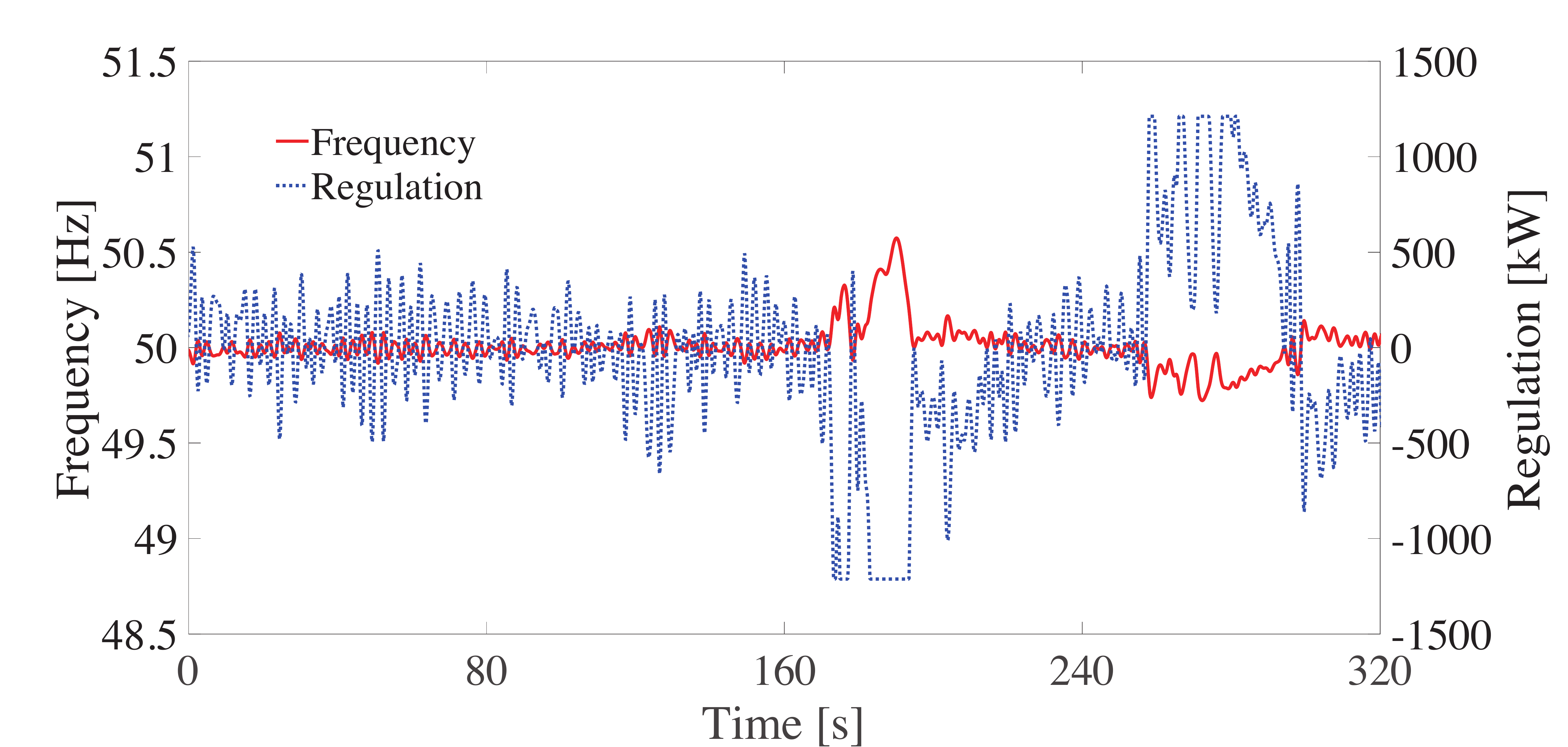} %{./figs/f_HIL_fully.eps}
\caption{%
Generated ancillary-service (AS) signal and associated frequency response when distributed electric vehicles can charge/discharge up to the maximum power without consideration of upper/lower limits.  
The result is derived by the Power-HIL simulation. 
}%
\label{fig:f_fully}
\end{figure}

\begin{figure}[t]
\centering
\includegraphics[width=.49\textwidth]{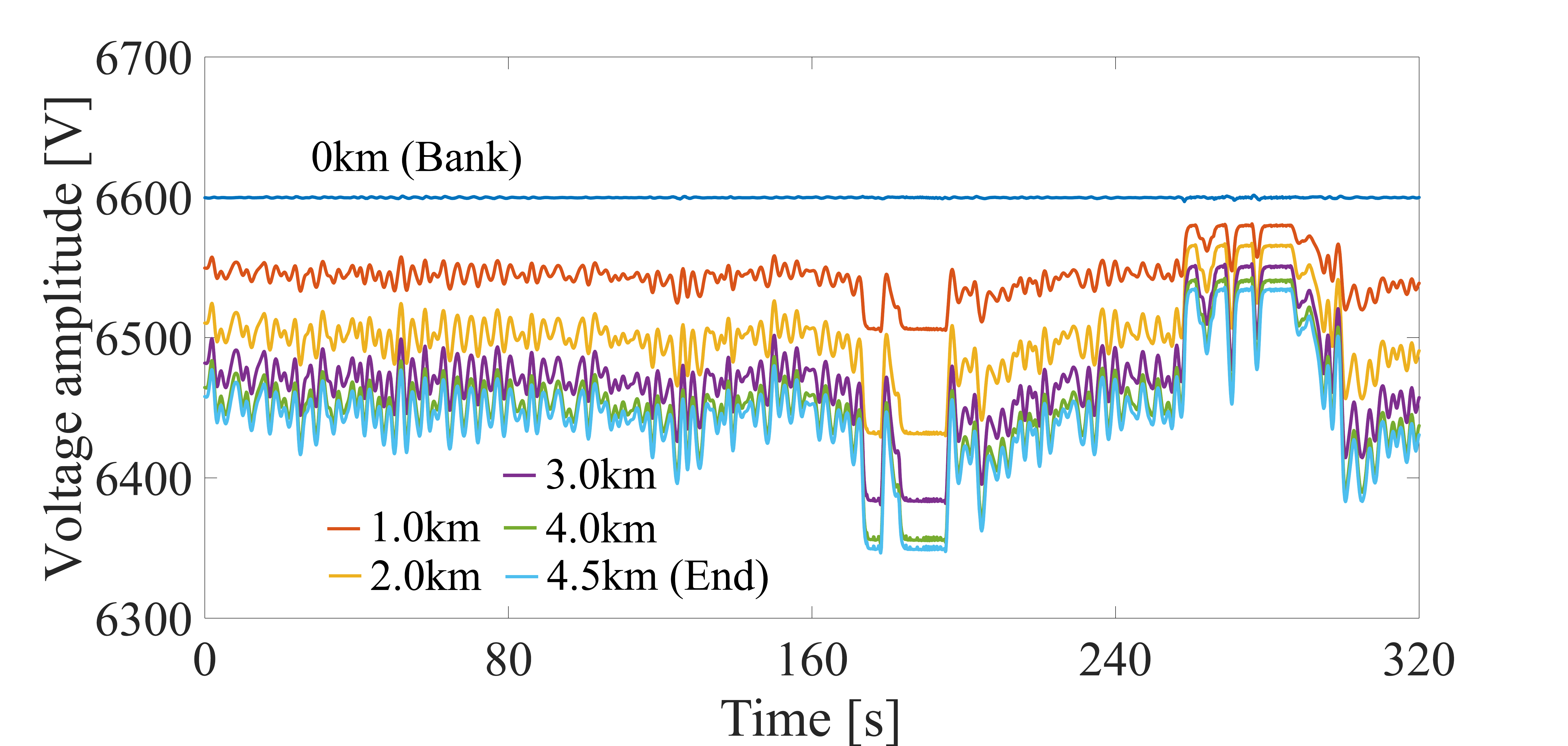} %{./figs/v_HIL_all.eps}
\includegraphics[width=.49\textwidth]{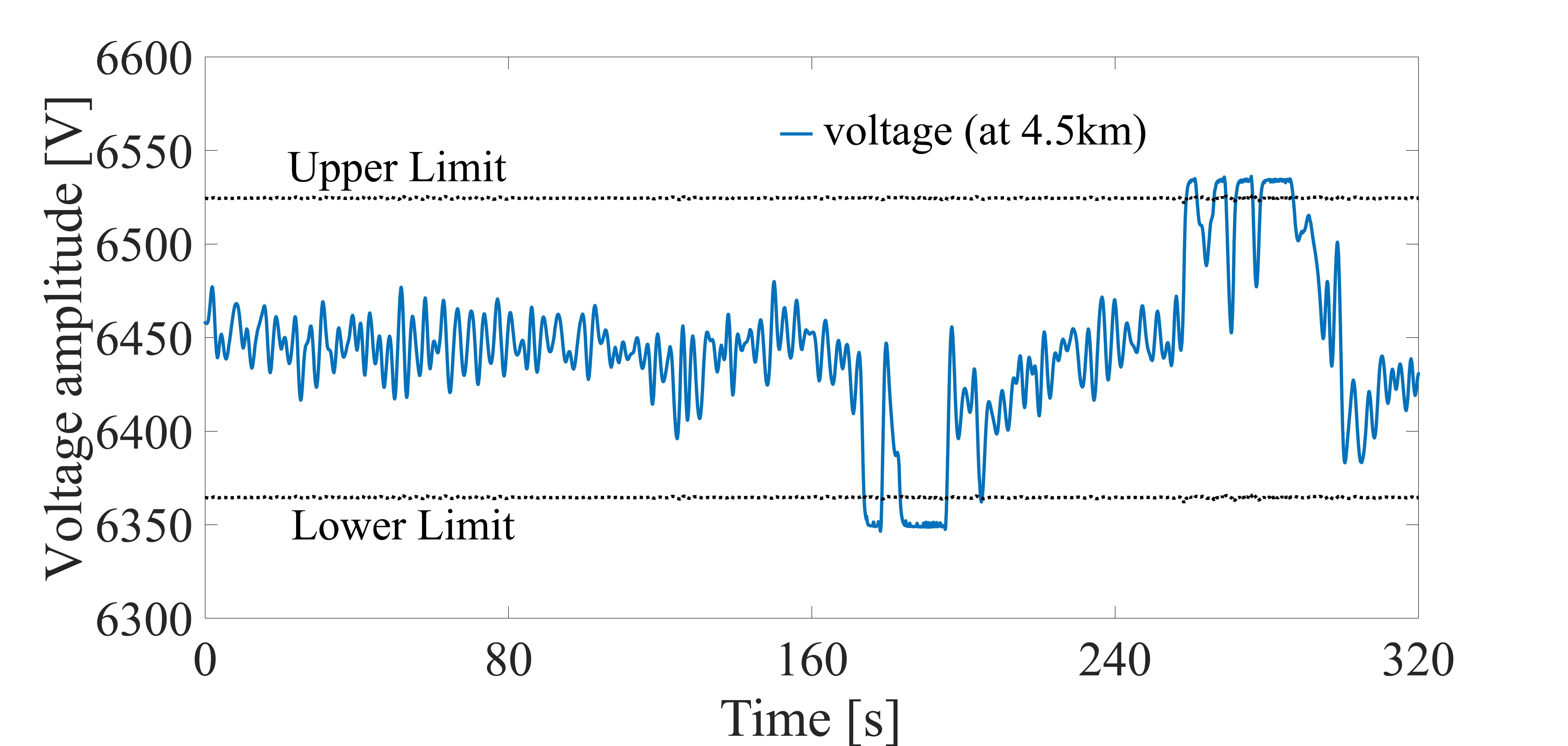} %{./figs/vL_HIL.eps}
\caption{%
Time series of sampled distribution voltage that are associated with Figure~\ref{fig:compare_hils}. 
(Upper) The voltage sampled at the six locations $x=0\,\U{km}$ (bank), $1.0\,\U{km}$, $2.0\,\U{km}$, $3.0\,\U{km}$, $4.0\,\U{km}$, and $4.5\,\U{km}$ (end) are plotted. 
(Lower) The voltage sampled at location $4.5\,\U{km}$ (end) is again plotted with the determined upper/lower bounds. 
}%
\label{fig:compare_v}
\end{figure}

First, we show the frequency dynamics under the design and see how the physical device affects the performance. 
Figure~\ref{fig:compare_hils} shows the Power-HIL simulation of the AS signal and associated frequency dynamics. 
The \emph{blue, dashed} line shows the time series of the AS signal for one feeder. 
The positiveness (or negativeness) of the AS signal implies the discharging (or charging) command. 
Here, we focus on the difference of time-delay and time-lag for the digital simulator and PCS. 
For this, we show in Figure~\ref{fig:compare_inout} the associated time series on input AS signal $P_\mathrm{EV}^\mathrm{ref}$ and output power $P_\mathrm{EV}$ in Figure~\ref{fig:delay} at Station~1 and Station~2. 
The \emph{red, solid} lines in Figure~\ref{fig:compare_inout} represent the AS signals, and the \emph{blue, dashed} lines do the output power. 
The two \emph{black, dashed} lines show the upper and lower limits as $\{-\alpha_\mathrm{cha} P_{\mathrm{EVs},i}^\mathrm{max} : i=1,\ldots,N_\mathrm{sta}\}$ and $\{\alpha_\mathrm{discha} P_{\mathrm{EVs},i}^\mathrm{max} : i=1,\ldots,N_\mathrm{sta}\}$. 
The upper/lower limits in Figure~\ref{fig:compare_inout} change once at the onset when the operation data on EVs at the four charging stations exhibit the stepwise change in Figure~\ref{fig:Number} (although not clearly shown in the left figures of Figure~\ref{fig:compare_inout} at about 120\,s). 
This is because $dV_\mathrm{cha, limit}$ and $dV_\mathrm{discha, limit}$ are fixed during the whole simulation, and the parameters $\alpha_\mathrm{cha}$ and $\alpha_\mathrm{discha}$ depend on the number of EVs only. 
In Figure~\ref{fig:compare_inout}(a) for Station~1 as PCS, we see that the time series of input AS signal and output power are not the same. 
This difference is manly due to the time-delay of ``Controller" in Figure~\ref{fig:Overview} and conversion characteristics of the real PCS. 
In Figure~\ref{fig:compare_inout}(b) for Station~2 on the digital simulator, the time series of input AS signal and output power are also not the same because the digital simulator considers time-delay and time-lag as shown in Figure~\ref{fig:delay}. 
The other two stations (Station~3 and 4) show qualitatively the same input/output responses as those in the figure (b). 
From these, we conclude that the quantitatively similar latency to the real one is simulated in the software-based stations.

Finally, we validate the proposed autonomous V2G design for the provision of PFC reserve. 
The frequency response in Figure~\ref{fig:compare_hils} is kept close to the nominal $50\,\U{Hz}$, and thus the PFC reserve by EVs works effectively for the frequency control. 
For comparison, we conduct the Power-HIL simulation in a case that EVs are fully distributed and can charge or discharge with their maximum power as described in Section~\ref{sec:exp}. Figure~\ref{fig:f_fully} shows the Power-HIL simulation of the AS signal and associated frequency response. 
In this case, because distributed EVs at each station can charge or discharge for the frequency control only, the grid's frequency approaches to the nominal ($50\,\U{Hz}$) better by comparison with Figure~\ref{fig:compare_hils}: see $[170\,\U{s}, 200\,\U{s}]$ in Figures~\ref{fig:compare_hils} and \ref{fig:f_fully}. 
This is expected in the design stage and indeed shown numerically in Section~\ref{sec:exp}. 
From the above simulations, we experimentally show that the autonomous V2G design works for the provision of PFC reserve, namely, the frequency control of the transmission grid.  

%%%
\subsection{Distribution Voltage Regulation}

Figure~\ref{fig:compare_v} shows the distribution voltage for the Power-HIL simulation associated with Figures~\ref{fig:compare_hils} and \ref{fig:compare_inout}. 
In this figure, the voltage sampled at the six locations including the end ($x=4.5\,\U{km}$) is plotted. 
The temporal deviations of voltage occur due to the AS signal to the four stations, the power consumption by the five loads, and the electric characteristics of the feeder. 

\begin{figure*}[t]
\centering
\begin{minipage}{.49\textwidth}
\centering
\includegraphics[width=\hsize]{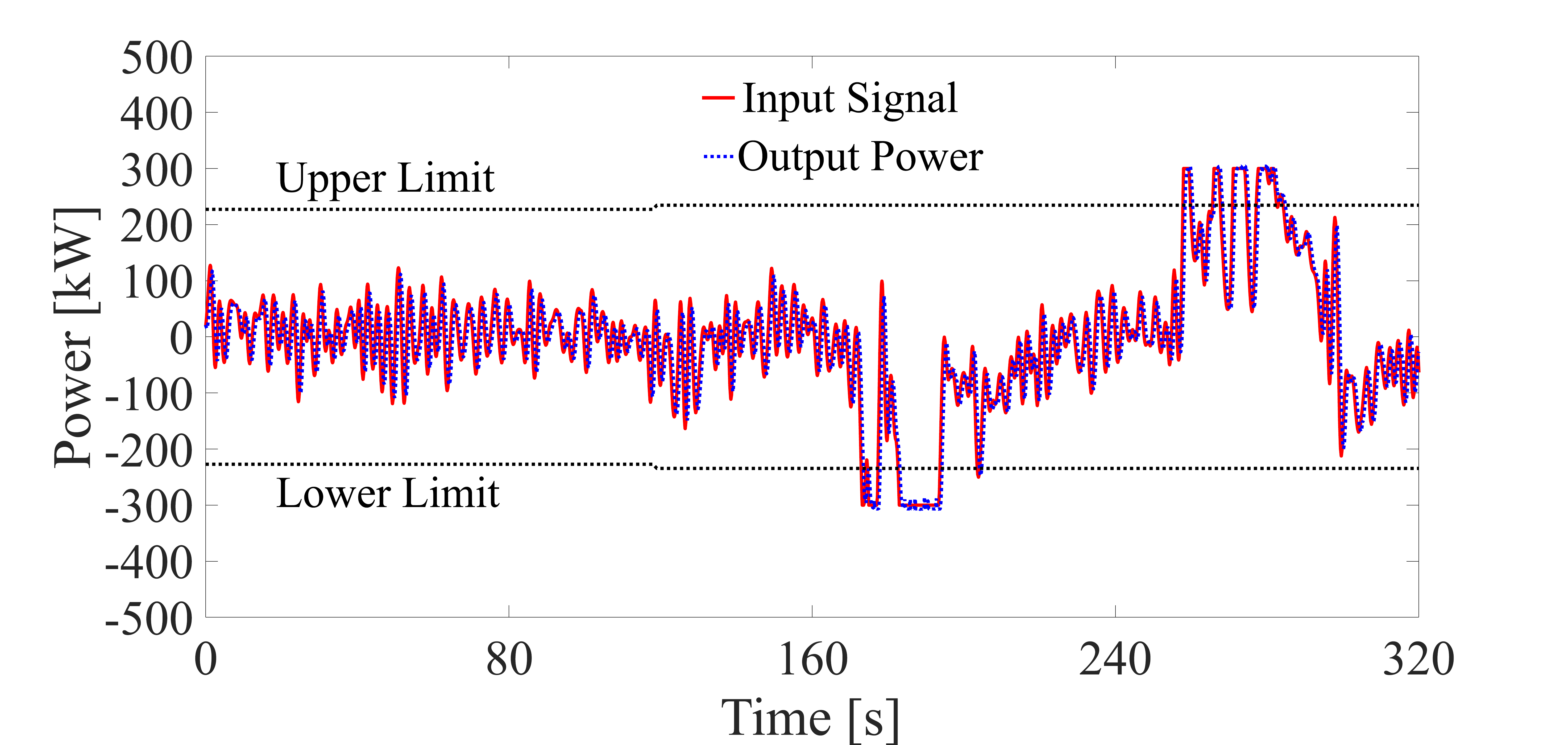} %{./figs/Sta11_limit.eps}
\subcaption{%
\footnotesize
Station~1
}%
\end{minipage}
\begin{minipage}{.49\textwidth}
\centering
\includegraphics[width=\hsize]{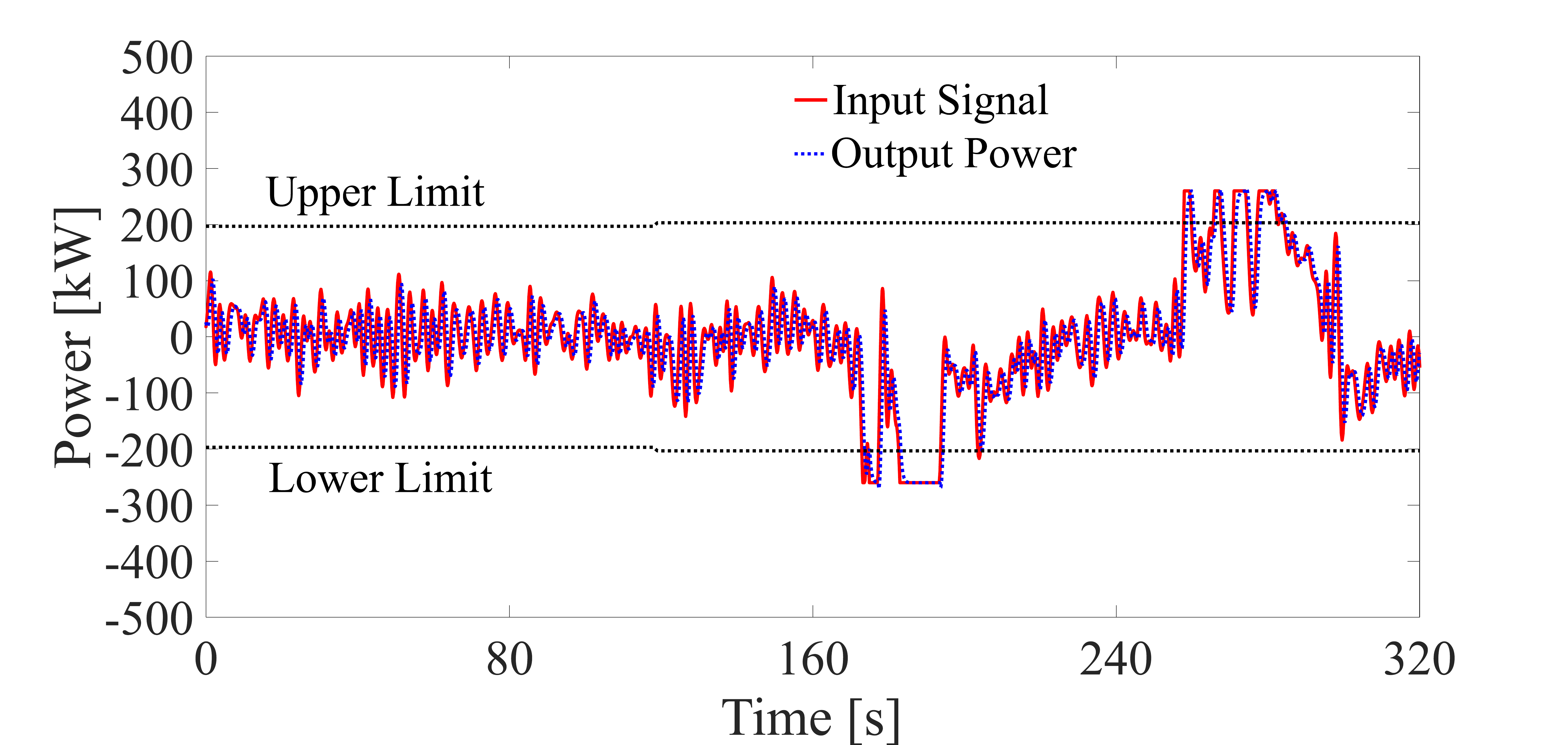} %{./figs/Sta22_limit.eps}
\subcaption{%
\footnotesize
Station~2
}%
\end{minipage}
\caption{%
Input/output responses of the two charging stations when distributed electric vehicles can charge/discharge up to the maximum power \emph{without} consideration of upper limit as in Figure~\ref{fig:f_fully}. 
The \emph{red, solid} lines show the input signals, the \emph{blue, dashed} lines show the output power, and the \emph{black, dashed} lines show the determined (but not used here) limits for the voltage regulation. 
}%
\label{fig:compare_inout2}
\end{figure*}

\begin{figure*}[t]
\centering
\begin{minipage}{.49\textwidth}
\centering
\includegraphics[width=\hsize]{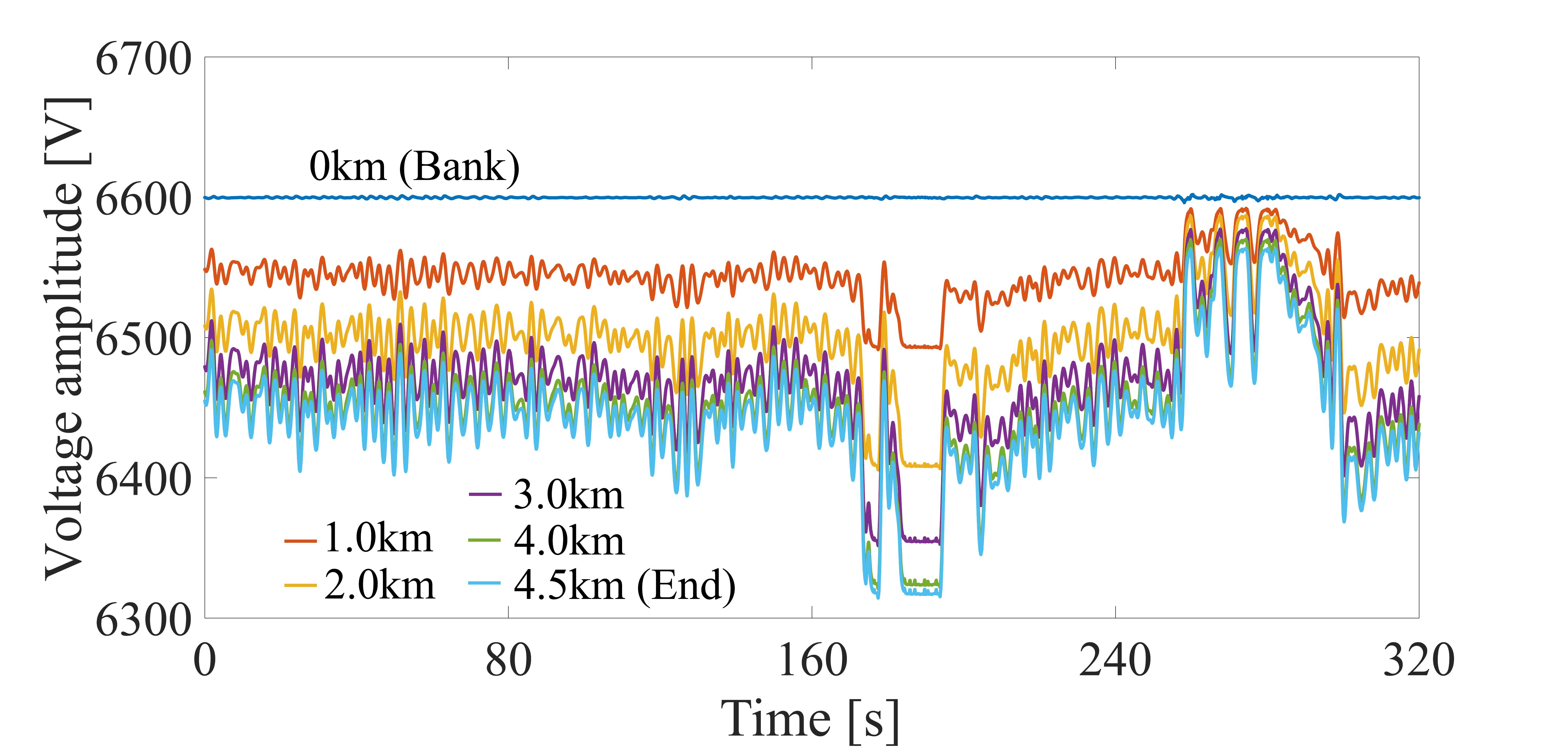} %{./figs/v_fully.eps}
\subcaption{%
\footnotesize
6 locations including end ($x=4.5\,\U{km}$)
}%
\end{minipage}
\begin{minipage}{.49\textwidth}
\centering
\includegraphics[width=\hsize]{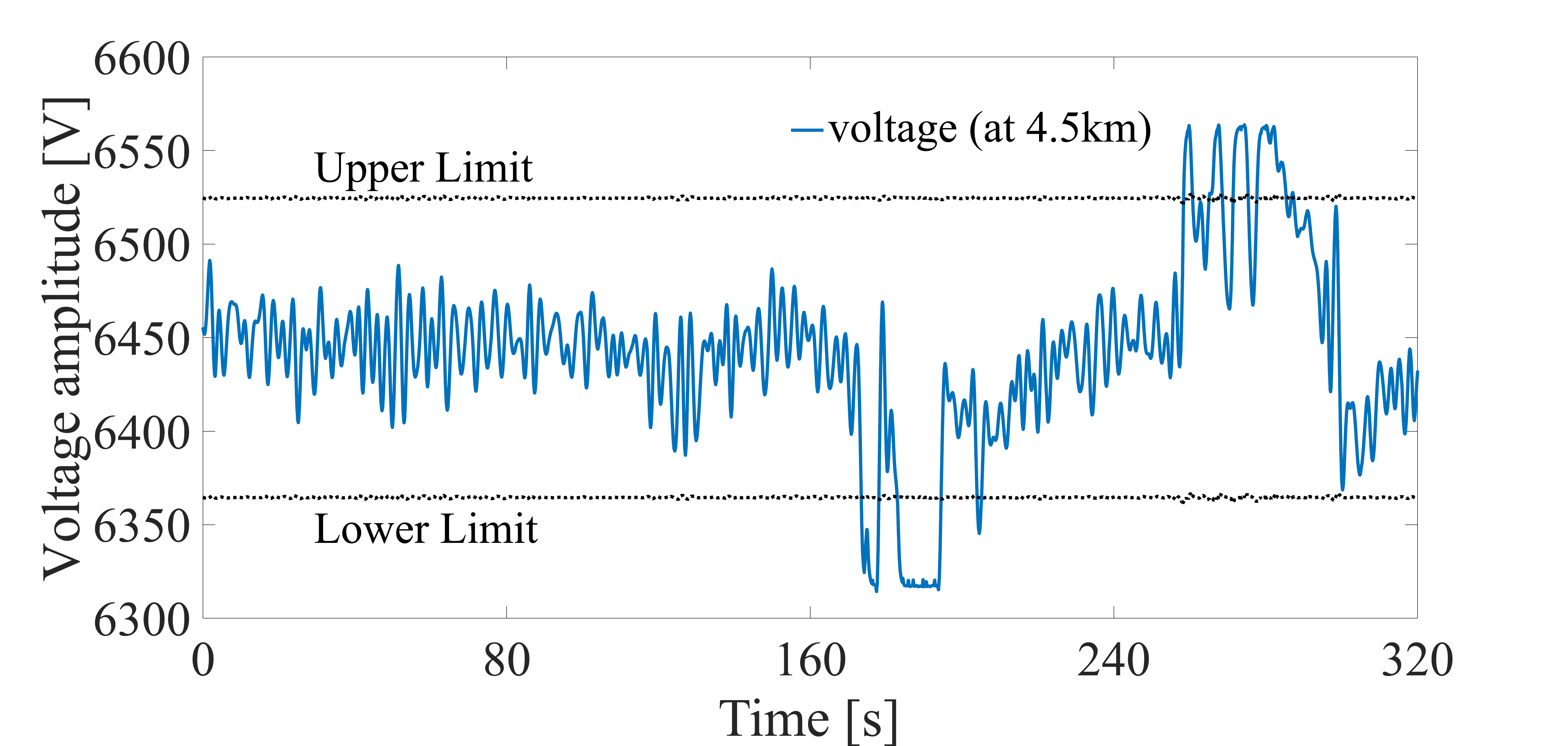} %{./figs/vL_fully.eps}
\subcaption{%
\footnotesize
End ($x=4.5\,\U{km}$) with upper/lower limits
}%
\end{minipage}
\caption{%
Time series of sampled distribution voltage associated with Figure~\ref{fig:compare_inout2}. 
}%
\label{fig:v_fully}
\end{figure*}

As the last task in this paper, we validate the proposed autonomous V2G design for the voltage regulation. 
In the same manner as Section~\ref{sec:PFCexp}, we consider the case in which distributed EVs can charge or discharge up to their maximum power \emph{without} consideration of upper limit. 
Figure~\ref{fig:compare_inout2} presents the input/output responses of the two charging stations in this case. 
The \emph{red, solid} lines in Figure~\ref{fig:compare_inout2} show the AS signals, and the \emph{blue, dashed} lines do the output power. 
The two \emph{black, dashed} lines show the upper and lower limits as $\{-\alpha_\mathrm{cha} P_{\mathrm{EVs},i}^\mathrm{max} : i=1,\ldots,N_\mathrm{sta}\}$ and $\{\alpha_\mathrm{discha} P_{\mathrm{EVs},i}^\mathrm{max} : i=1,\ldots,N_\mathrm{sta}\}$. 
Unlike Figure~\ref{fig:compare_inout} with the consideration of upper limit, we see that the magnitude of output power goes over the limits in Figure~\ref{fig:compare_inout2}: see the responses during $[170\,\U{s},200\,\U{s}]$ and $[260\,\U{s},280\,\U{s}]$. 
The associated temporal deviations of voltage are shown in Figure~\ref{fig:v_fully}. 
Here, the charging/discharging power at two charging stations is not kept within the determined range for the voltage regulation in Figure~\ref{fig:compare_inout2}. 
We see that the voltage of Figure~\ref{fig:v_fully}(b) is largely deviated from the upper/lower limits compared to the bottom in Figure~\ref{fig:compare_v}. 
This implies that the autonomous V2G design using the upper/lower limits is effective for reducing the deviation of distribution voltage at the end. 
It should be mentioned that the voltage in Figure~\ref{fig:compare_v} is over the lower limit because of the intrinsic approximation error for the computation of upper bound. 
This is studied substantially in Section~\ref{sec:exp} and shows that it can be reduced by appropriately choosing the grid's condition. 
As suggested in Figure~\ref{fig:error}, it would be effective to decrease the amount of the loading condition for reducing the degree of violation. 
The Power-HIL simulation shows that the autonomous V2G design also works for the voltage regulation.

%%%%
%%%%
\section{Conclusion}
\label{sec:conclusion}

This paper proposed the new design of autonomous V2G for the provision of multi-objective AS. 
The design is computationally simple (no need of optimization), easily implemented (as demonstrated in Section~\ref{sec:HIL}), and relevant 
because it is guided by the physics-based models of power grids and the solid mathematical approach (as shown in Sections~\ref{sec:ODE}). 
Then, we numerically show that the proposed design effectively works for not only the provision of PFC reserve and the regulation of distribution voltage. 
It is also experimentally shown that the design works in a practical situation where inevitable latency is involved, which shows the practical feasibility of the design.  

This study, which built upon on the recent development of coupling of transportation and energy systems, is another step toward this development. 
Several research directions are possible. 
For utilization of the proposed design in practical distribution grids, conventional metrics of their operation such as significant losses should be included. 
In this paper, we considered the control by DSO, and we did not consider the operational strategy of EV-sharing operator, e.g., reported in \cite{Kawashima:CCTA17}, for assignment and reallocation of EVs. 
It is in our future work to explore a detailed design of the architecture in Section~\ref{sec:framework} by taking both the operators into account. 
It is also important to develop a platform for the architecture in terms of communication, data integration, and cybersecurity. 
In addition to these technological studies, it is necessary to conduct the so-called economic evaluation of the architecture. 
A social benefit resulting from the architecture will be clarified through the evaluation. 
By combining the technological and economical studies, it is expected to establish explicit guidelines for the successful implementation of the proposed architecture. 
In this, specific requirements on amount of reserve and its ramping capability to be remunerated as an AS should be clarified.

%%%%%
\section*{Acknowledgement}

The authors appreciate the valuable suggestions of the reviewers while preparing the manuscript.
This work was supported in part by Japan Science and Technology Agency, Core Research for Evolutional Science and Technology (JST-CREST) Program [grant number JP-MJCR15K3].

%%%%
%%%%
%!TEX root = main_v0.tex
%%%%%
%%%%%

\end{document}